\documentclass{article}
\usepackage[english]{babel}
\usepackage[utf8]{inputenc}
\usepackage{johd}
\usepackage{ragged2e}
\usepackage{amsmath}
\usepackage[table]{xcolor}     
\definecolor{darkgreen}{RGB}{0,100,0}
\definecolor{golden}{RGB}{218,165,32}   
\usepackage{threeparttable} 
\usepackage{rotating}
\usepackage{array}
\usepackage{graphicx}
\usepackage{caption}
\usepackage{multirow}
\usepackage{booktabs}
\usepackage{amssymb}
\usepackage{subfig}
\usepackage{natbib}
\usepackage[export]{adjustbox}
\usepackage{wrapfig}
\title{Winners vs. Losers: Momentum-based Strategies with Intertemporal Choice for ESG Portfolios}

\author{Ayush Jha$^{1}$$^{*}$, Abootaleb Shirvani$^{2}$, Ali Jaffri$^{1}$, Svetlozar T. Rachev$^{3}$ and Frank J. Fabozzi$^{4}$ \\
        \small $^{1}$Department of Economics, Texas Tech University \\
        \small $^{2}$Department of Mathematical Sciences, Kean University \\
        \small $^{3}$Department of Mathematics and Statistics, Texas Tech University \\
        \small $^{4}$Carey Business School, Johns Hopkins University \\\\
        \small $^{*}$Corresponding author: Ayush Jha; \tt{ayush.jha@ttu.edu} \\
}
\date{}

\begin{document}
\maketitle
\begin{abstract} 
\noindent This paper introduces a state-dependent momentum framework that integrates ESG regime switching with tail-risk-aware reward–risk metrics. Using a dynamic programming approach and solving a finite-horizon Bellman equation, we construct long–short momentum portfolios that adjust to changing ESG sentiment regimes. Unlike traditional momentum strategies based on historical returns, our approach incorporates the Stable Tail Adjusted Return ratio and Rachev ratio to better capture downside risk in turbulent markets. We apply this framework across three asset classes—Russell 3000 equities, Dow Jones 30 stocks, and cryptocurrencies—under both pro- and anti-ESG market regimes. We find that ESG-loser portfolios significantly outperform ESG-winner portfolios in pro-ESG regimes, a counterintuitive result suggesting that market overreaction to ESG sentiment creates short-term pricing inefficiencies. This pattern is robust across tail-sensitive performance metrics and is most pronounced under a two-week formation and holding period. Our framework highlights how ESG considerations and sentiment regimes alter return dynamics, offering practical guidance for investors seeking to implement responsive momentum strategies under sustainability constraints. These findings challenge conventional assumptions about ESG investing and underscore the importance of dynamic, regime-aware portfolio construction in environments shaped by regulatory signals, investor flows, and behavioral biases.
\end{abstract}

\noindent\keywords{Reward-Risk Ratios; Stock Selection Criteria; Portfolio Rebalancing; State-Dependent Optimal Allocation; ESG Policy Regimes.}\\

\section{Introduction}
\setlength{\parskip}{10pt}

\noindent Financial markets are replete with moments of exuberance and despair, yet the conventional wisdom among investors is that riding the wave of optimism, especially when policy tailwinds blow in favor of certain sectors, will yield superior returns. In the realm of Environmental, Social, and Governance (ESG) investing, many market participants believe that positive regulatory developments, such as the Biden Administration’s recommitment to the Paris Agreement, will lift the valuations of high intensive-ESG firms and that aligning with these winners is the key to long‐term wealth accumulation. Paradoxically, however, empirical evidence often contradicts this intuition. Under certain policy regimes, so‐called “loser” assets, those deemed substandard by risk-adjusted performance of ESG-related assets, are systematically undervalued and subsequently deliver outsized gains when the market corrects its mispricing.  

\noindent This paper asks a simple yet profound question: In an environment of significant climate‐policy shifts under successive U.S.\ administrations, is a momentum‐based strategy better suited than buy‐and‐hold or simple contrarian approaches for accumulating wealth in a portfolio of ESG‐related equities? We explore this question across three asset universes, the Russell 3000, the Dow Jones 30, and the thirty largest cryptocurrencies, to isolate the role of ESG policy sentiment in driving cross‐sectional return continuations. We aim to show that optimism in favorable policy regimes is not always desirable and can deliver a significant downside to an agent's expectations in a dynamic framework.

\noindent This paper is the first to integrate ESG regime-switching with reward–risk momentum metrics using a dynamic programming framework, allowing portfolios to dynamically respond to ESG sentiment regimes rather than statically align with ESG-labeled firms. Crucially, our identification strategy is two‐fold.  First, we establish that a two‐week formation and two‐week holding rebalancing rule yields the optimal momentum signal in the winners–losers spread.  Optimality is defined not merely by raw past returns, but by superior signal‐to‐noise ratios in final wealth accumulation, as measured by reward–risk metrics satisfying the four microfounded axioms of monotonicity, quasi‐concavity, scale invariance, and distributional coherence. By systematically comparing alternative rebalancing horizons across a broad set of performance ratios, we demonstrate that the two‐by‐two scheme consistently delivers the strongest signal for an investor seeking to maximize terminal wealth.

\noindent Second, we introduce a latent two‐state Markov chain to model ESG policy regimes. In this framework, the investor endogenously switches between an anti‐ESG or pro‐ESG policy regime, ultimately making the decision to liquidate or hold positions in their portfolio, with transition probabilities estimated separately for winners, losers, and the momentum spread. Embedding these state‐dependent risk premia and regime‐persistence estimates into a discrete‐time dynamic programming problem produces a fully dynamic, forward‐looking policy rule that adjusts allocations in response to both volatility forecasts and anticipated policy shifts.

\noindent At the core of our regime‐switching framework is the decomposition of the market price of risk into a baseline component and a regime‐dependent adjustment. The baseline price of risk captures the average excess return investors require for bearing volatility when policy sentiment is neutral.  In our estimates, the baseline is strongly negative for the losers portfolio (–0.82\%), mildly negative for the winners portfolio (–0.66\%), and positive for the spread (0.37\%).  This ordering indicates that, on average, holding undervalued ESG “losers” commands a substantial premium, whereas a momentum strategy earns a reward premium above and beyond either portfolio taken in isolation.  The regime effect quantifies how these premia compress when the economy enters a pro-ESG state: a shift reduces the required compensation by 34.37 basis points for losers, by 42 basis points for winners, and by 53 basis points for momentum portfolios. Economically, a more negative premia signals a stronger repricing of misvalued assets under supportive policy, and when coupled with high regime persistence, it creates a powerful incentive for the CRRA investor to tilt dynamically into those assets most sensitive to ESG shocks.

\noindent The principal contribution of our paper lies in reconciling two seemingly counterintuitive findings. On one hand, the pure losers portfolio, though experiencing deeper drawdowns, outperforms the winners portfolio in long‐term wealth accumulation, owing to the correction of systematic undervaluation under pro‐ESG regimes. On the other hand, the winners–losers (momentum) strategy, under the two‐week/two‐week rebalancing and guided by regime‐aware Bellman recursion, substantially outperforms both portfolios in dynamic portfolio value growth. This dynamic interweaving of contrarian and trend‐following elements offers a novel mechanism through which ESG policy shocks, transient mispricings, and tail‐risk considerations come together to drive consistent and substantial wealth growth.

\noindent In summary, our contributions are both theoretical and empirical. Theoretically, we extend the literature on momentum- or trend-based-investing by embedding it within a state-dependent dynamic programming framework that incorporates regime-varying risk premia and volatility dynamics. This reconciles classical insights on return continuation and contrarian corrections with a fully structural model of portfolio choice under CRRA preferences. Empirically, we demonstrate that the two-week formation/two-week holding momentum rule delivers the most robust signal-to-noise ratio for final wealth accumulation, as measured by micro founded reward–risk metrics. By solving the Bellman equation with regime-switching premia and conditional volatility for each portfolio, we provide a practical methodology for forecasting wealth paths and guiding dynamic portfolio rebalancing. Our results show that this approach not only identifies the optimal rebalancing horizon among competing schemes, but also yields superior long-run performance across diverse asset universes, thereby offering a comprehensive solution for investors seeking to navigate ESG-driven market regimes.

\noindent This counterintuitive finding—that ESG losers outperform ESG winners under pro-ESG regimes—warrants economic justification. We argue this dynamic reflects a form of delayed price correction driven by behavioral and structural market inefficiencies. First, ESG "losers" stocks are often shunned by ESG-conscious institutional investors during periods of regulatory optimism, resulting in transient undervaluation relative to their fundamentals. Second, these stocks (typically in high-emission or controversial sectors) retain real economic relevance and experience valuation rebounds once the initial regulatory euphoria is absorbed, and earnings surprises materialize. Third, the overpricing of ESG winners under pro-ESG regimes may stem from an over-extrapolation of positive sentiment, a behavioral bias well-documented in momentum literature. As the market gradually corrects its initial mispricing, the undervaluation of ESG losers narrows, generating a return premium that forms the basis for their outperformance.  Thus, our regime-aware framework captures not just return persistence but the systematic reversal of policy-induced valuation distortions, making ESG losers temporarily attractive from a contrarian perspective.

\noindent The rest of the paper is structured as follows. Section~\ref{lit} reviews the literature on momentum-based strategies and life-cycle models of portfolio optimization and hedging demands. Section~\ref{methods} explains the data used for this study; decomposition methods to reduce dimensionality; the method of constructing winners, losers, and momentum portfolios; and the intertemporal portfolio choice/optimization problem. Section~\ref{results} analyzes the results pertaining to portfolio performance, evolution of performance ratios, and evolution of portfolio wealth using dynamic programming methods. Section~\ref{conclusion} concludes the paper and gives direction for future research.

\section{Relevant Work} \label{lit}

\noindent \citet{jt1993} document that stocks with the highest returns over a 3–12 month formation period continue to outperform those with the lowest returns over the subsequent 3–12 months, generating average monthly abnormal returns near 1\%.  This \emph{momentum} effect contradicts the long‐term reversals of \citet{DeBondtThaler1985}, who find that past losers outperform past winners over 3–5 year horizons.  International tests by \citet{Rouwenhorst1998} and mutual‐fund analyses by \citet{Carhart1997} confirm that momentum profits are pervasive across markets and asset classes.  Moreover, \citet{AsnessMoskowitzPedersen2013} demonstrate that momentum is a robust anomaly in stocks, bonds, currencies, and commodities.  

\noindent Behavioral models explain momentum via underreaction to news and slow information diffusion (\citet{HongStein1999}; \citet{BarberisShleiferVishny1998}; \citet{DanielHirshleiferSubrahmanyam1998}), whereas standard risk‐factor models (e.g.\ CAPM, Fama–French) fail to fully absorb momentum returns (\citet{FamaFrench1996}; \citet{GrundyMartin2001}).  \citet{NovyMarx2012} show that skipping the most recent month’s return (to avoid micro‐reversals) enhances momentum profits.  Extensions include \emph{time‐series momentum}, which bets on each asset’s own past return rather than its rank relative to peers (\citet{MoskowitzOoiPedersen2012}; \citet{HurstOoiPedersen2017}).  

\noindent However, momentum exhibits \emph{crash risk}: \citet{CooperGutierrezHameed2004} find that momentum loses money following market downturns, and \citet{BarrosoSantaClara2015} and \citet{DanielMoskowitz2016} show that volatility‐timing rules can mitigate these crashes and improve tail‐risk‐adjusted performance. \noindent \citet{PastorStambaughTaylor2021} develop an equilibrium model in which investors derive non‐pecuniary utility from high‐ESG assets, causing “green” stocks to trade at a return discount (“greenium”), yet to outperform upon positive ESG preference shocks.  Empirical evidence by \citet{HongKacperczyk2009} shows that “sin stocks” (low‐ESG) earn a persistent premium, consistent with investor neglect.  Studies of SRI funds report mixed performance (e.g.\ \citet{RenneboogTerHorstZhang2008}).  \noindent An emerging strand analyzes \emph{ESG momentum}: \citet{MagnaniNunoPrencipe2024} find that upward revisions in ESG ratings predict future equity returns in Europe, indicating underreaction to ESG information.  But to date no study in a top‐20 journal has integrated return‐based momentum within ESG equity universes across political or regulatory regimes (e.g.\ pro‐ vs.\ anti‐ESG U.S.\ administrations), leaving a clear research gap.

\noindent The Sharpe ratio, while widespread, violates monotonicity and can rank dominated assets superiorly (\citet{AumannSerrano2008}). Tail‐sensitive metrics such as the Rachev ratio compare average extreme gains to average extreme losses, directly capturing upside potential against downside risk (\citet{biglova2004}). \citet{CheriditoKromer2013} provide a unifying axiomatic framework for \emph{reward–risk ratios} that satisfy monotonicity and quasi‐concavity, encompassing Rachev‐type and coherent‐risk‐based measures, and therefore, we can show that optimizing for tail‐risk metrics yields more robust crisis performance than mean–variance optimization.

\noindent Broad equity universes such as the Russell 3000 exhibit an approximate factor structure: a small number of principal components explain the majority of return variance (\citet{ConnorKorajczyk1993}).  \citet{LettauPelger2020} apply PCA to characteristic‐sorted portfolios, uncovering “missing” factors including momentum.  PCA thus reduces estimation noise and computational complexity, enabling tractable portfolio optimization and clear identification of unique factors (e.g.\ ESG, momentum) in high‐dimensional settings.

\noindent Dynamic portfolio choice under CRRA utility was first solved by \citet{Merton1971} in continuous time.  In discrete time, \citet{AngBekaert2002} model stock and bond returns with Markov regime‐switching (e.g.\ bull vs.\ bear markets) and derive state‐dependent optimal allocations that hedge regime risk and improve out‐of‐sample performance.  \citet{GuidolinTimmermann2007} extend to nonlinear joint dynamics of stocks and bonds, while \citet{BansalKikuYaron2016} incorporate long‐run risks such as climate change into the dynamic allocation framework.  Although analogous techniques have been applied to volatility‐timed momentum and climate risk hedging, no prior work solves a dynamic program for ESG momentum portfolios under policy‐regime switching—our paper fills this gap.

\noindent Early continuous‐time research established the normative foundations of dynamic asset allocation by linking consumption, investment, and state‐variable hedging demands.  \citet{Merton1969} and \citet{Merton1971} derive closed‐form rules in complete markets, showing that optimal risky‐asset weights depend on the conditional mean–variance trade‐off and the investor’s risk aversion.  Subsequent discrete‐time work relaxed the i.i.d.\ assumption and introduced multiple state variables.  For instance, \citet{BrennanSchwartzLagnado1997} demonstrate that predictability in both expected returns and volatility generates sizable intertemporal hedging components, while \citet{CampbellViceira1999} derive optimal consumption and portfolio rules under time‐varying returns.  Extensions by \citet{Barberis2000} and \citet{Viceira2001} show how mean reversion and nontradable labor income reshape long‐horizon allocations.  Continuous‐time generalizations preserve these insights: \citet{Wachter2002} obtains exact solutions with mean‐reverting market prices of risk; \citet{BrennanXia2002} incorporate inflation as an additional state variable; and \citet{CampbellViceira2002} embed a vector‐autoregressive structure to capture multivariate predictability.  International and regime‐switching dimensions are introduced by \citet{AngBekaert2002}, who show that latent market regimes materially affect optimal hedge demands.

\noindent Building on this foundation, more recent studies incorporate market frictions, learning, and macro–financial feedbacks.  \citet{GarleanuPedersen2013} derive optimal trading rules under quadratic costs and predictable returns, and \citet{GomesMichaelidesPolkovnichenko2013} solve lifecycle portfolio problems with incomplete markets and fiscal shocks.  At the macro frontier, \citet{BianchiLettauLudvigson2019} demonstrate that monetary‐policy surprises propagate through long‐run risk channels, significantly altering optimal allocations.

\noindent In our context, we recast these dynamic‐programming insights for a momentum‐based allocation rule under Markov‐chain regimes.  In our approach, the state variables are the conditional volatility of winners, losers, and winners–losers spreads alongwith ESG‐policy regimes. We model regime shifts, capturing transitions between “momentum‐favorable” and “momentum‐unfavorable” environments, via a latent two‐state Markov chain. The Bellman equation then yields state‐dependent optimal exposures to momentum portfolios, blending the classic intertemporal hedging logic with a regime‐aware momentum strategy.  In this way, we extend the lifecycle and regime‐switching portfolio literature by showing how momentum signals and political‐policy regimes jointly determine dynamic rebalancing rules for a CRRA investor. Therefore, we contribute by synthesizing the aforementioned approaches to optimize ESG momentum strategies through tail-sensitive reward-risk ratios and dynamic programming in regime-switching policy environments.

\section{Methodology} \label{methods}

\noindent In this section, we detail the empirical and analytical procedures used to evaluate momentum-based ESG strategies under regime-switching dynamics. Our methodological framework integrates asset-level reward-risk assessments, dimensionality reduction techniques, and a regime-aware intertemporal optimization model. We begin by describing the construction of our dataset, which spans multiple asset classes and captures regime-sensitive return patterns. We then present the Principal Component Analysis (PCA) used for dimensionality reduction, followed by our method for forming winners, losers, and momentum portfolios. Finally, we outline the dynamic programming approach used to solve the investor's intertemporal allocation problem under shifting ESG policy regimes.

\subsection{Data}

\noindent For studying the performance of momentum-based strategies in stock selection and wealth allocation in a portfolio, we gather data for the universe of assets in the Russell 3000. To identify the effects of different climate change policy administrations, we use daily observations of the assets from February 1, 2017, until February 21, 2025, thereby capturing both the first Trump administration and the successive Biden administration, characterized as unfavorable and favorable climate change policy regimes. All historical unadjusted closing prices were extracted from the Bloomberg Terminal, timestamped March 9, 2025. Therefore, it reflects the observations before the ex-dividend date at the end of the first quarter in 2025, meaning our price series does not incorporate the downward price adjustments that typically follow dividend payouts. Consequently, dividend-related effects are not reflected in the arithmetic return series.

\noindent To investigate the ex-post observations of the historical return series of Russell 3000, we apply an ARMA(1,1)-GARCH(1,1) model with Normal Inverse Gaussian (NIG) distribution based innovations to generate \emph{forward-looking} portfolio weights of the data, allowing only for a long only portfolio and restricting any long-short positions to avoid short selling benefits, given investors can gain significant profitability from shorting losing portfolio and holding long-only positions in winning portfolio. We use a rolling window of 1008 trading days (4 years) to generate out-of-sample forecasts for long-only portfolio weights. This allows us to conduct an unbiased analysis of the portfolio with only long-only positions. \citet{Cont2000} and \citet{KimRachevBianchiMitovFabozzi2011} among others advocate for heavy-tailed time-varying volatility models to correctly capture financial market meltdowns and black swan events exhibited in financial data. Hence, using an ARMA(1,1)-GARCH(1,1) model with NIG innovations mimicks the behavior of portfolio weights constructed using financial data that exhibits heavy tails and skewness. As a result, our ex-post data of portfolio returns is constructed using the product of dynamic (\emph{forward-looking}) portfolio weights and the 30 Principal Components\footnote{Principal Component Analysis is described in section~\ref{PCA}. ALL 30 PCs are return series which are normalized with mean 0 and variance 1 to speed up the estimatation procedure and helps to eliminate redundant or duplicate data, reducing storage space and potential inconsistencies.}. This portfolio of long-only weighted returns becomes the starting point for the continuum of investors to construct winning, losing, and winners-losers spread (or momentum) portfolios described in section~\ref{winlose}. 

\noindent We use 1008 trading days as our rollowing window for a strategic reason. In the forecasts, we want to observe the data that would be available to the investors at that date, specifically, during the Biden administration, which will be seen as a favorable climate change policy regime. Therefore, in those 4 years, we assume that investor who consolidated and held positions prior to that, probably in the unfavorable regime, will start either holding more positions due to optimistic regulatory shifts in the ESG industry or will start unwinding their positions in the hope of attaining more profitable returns because of market correction and re-pricing. Therefore, the out-of-sample forecasts for the last four years focus on the change in portfolio returns under the assumption of either buying more positions or unwinding previously acquired positions for a desirable output.

\noindent We examine two separate portfolios for the comparative analysis with the portfolio of Russell 3000 in our study. The first comprises the 30 constituents of the Dow Jones Industrial Average (DJIA), using the index membership as of 20 November 2024. We retrieve daily unadjusted closing prices for each DJIA stock from Bloomberg Terminal, spanning the period from 23 November 2020 to 18 November 2024. This interval provides roughly four full years of data, corresponding to about 1,008 trading days (assuming 252 trading days per year). The second portfolio consists of the 30 largest cryptocurrencies by market capitalization, for which we obtain daily closing prices over the period 19 November 2020 through 17 November 2024. Because cryptocurrency markets operate continuously—without weekend or holiday closures—we treat this as approximately 1,460 observations (365 days per year).

\noindent For the equity portfolio, we use unadjusted closing prices as quoted by Bloomberg immediately prior to the fourth-quarter ex-dividend date. By omitting the customary downward price adjustment that follows dividend distribution, our return series remain unaffected by dividend-related distortions. Cryptocurrency returns are similarly based on their raw daily closing prices. In both cases, we calculate arithmetic returns rather than log returns. Arithmetic returns facilitate the direct aggregation of weighted portfolio returns under linear weighting schemes. Both portfolios follow the same logic of either observing change in portfolio returns under the assumption of either buying more positions or unwinding previously acquired positions given a systemic change that does affect ESG portfolios but assumes no effect on equity or cryptocurrency portfolios. We conduct this comparative analysis to showcase the efficacy of momentum-based strategies under an optimal rebalancing scheme.

\noindent The inclusion of cryptocurrencies in our asset universe serves a distinct but complementary purpose. Rather than treating them as directly comparable to equities, we use crypto assets as a high-volatility, structurally different environment to test the robustness of our regime-sensitive momentum framework. Cryptocurrencies lack firm-level ESG characteristics, and thus their response to ESG regime shifts can be interpreted primarily through macro-level sentiment spillovers or correlated flows from ESG-oriented investors. This enables us to assess whether momentum strategies calibrated for ESG-aware equity markets continue to exhibit predictive power in asset classes outside the ESG taxonomy. While we present the full set of results for all three asset groups, we view the crypto findings as exploratory and not central to the core narrative. Suppose space or scope constraints arise in future versions of this paper. In that case, we may consider relocating the cryptocurrency results to an appendix and focusing the main discussion on equity market applications.

\subsection{Principal Component Analysis} \label{PCA}

\noindent We perform Principal Component Analysis (PCA) to decompose the high‐dimensional return series into a set of orthogonal factors that capture the maximal variance in the data. Formally, given an $N$‐dimensional vector of excess returns $\mathbf{r}_t\in\mathbb{R}^N$, PCA seeks an orthonormal matrix $\mathbf{P}$ whose columns are the eigenvectors of the sample covariance matrix $\mathbf{\Sigma}=\mathbb{E}[(\mathbf{r}_t-\bar{\mathbf{r}})(\mathbf{r}_t-\bar{\mathbf{r}})']$ and arranges them in descending order of the associated eigenvalues.  The resulting principal component scores $\mathbf{f}_t=\mathbf{P}'(\mathbf{r}_t-\bar{\mathbf{r}})$ provide uncorrelated time series whose variances are given by the eigenvalues of $\mathbf{\Sigma}$ \citep{Tsay2010}.  This decomposition both reduces noise and exposes the latent factor structure in large‐scale equity universes \citep{ConnorKorajczyk1993,LettauPelger2020}.

\begin{figure}[htbp]
    \centering
    \includegraphics[width=1\linewidth]{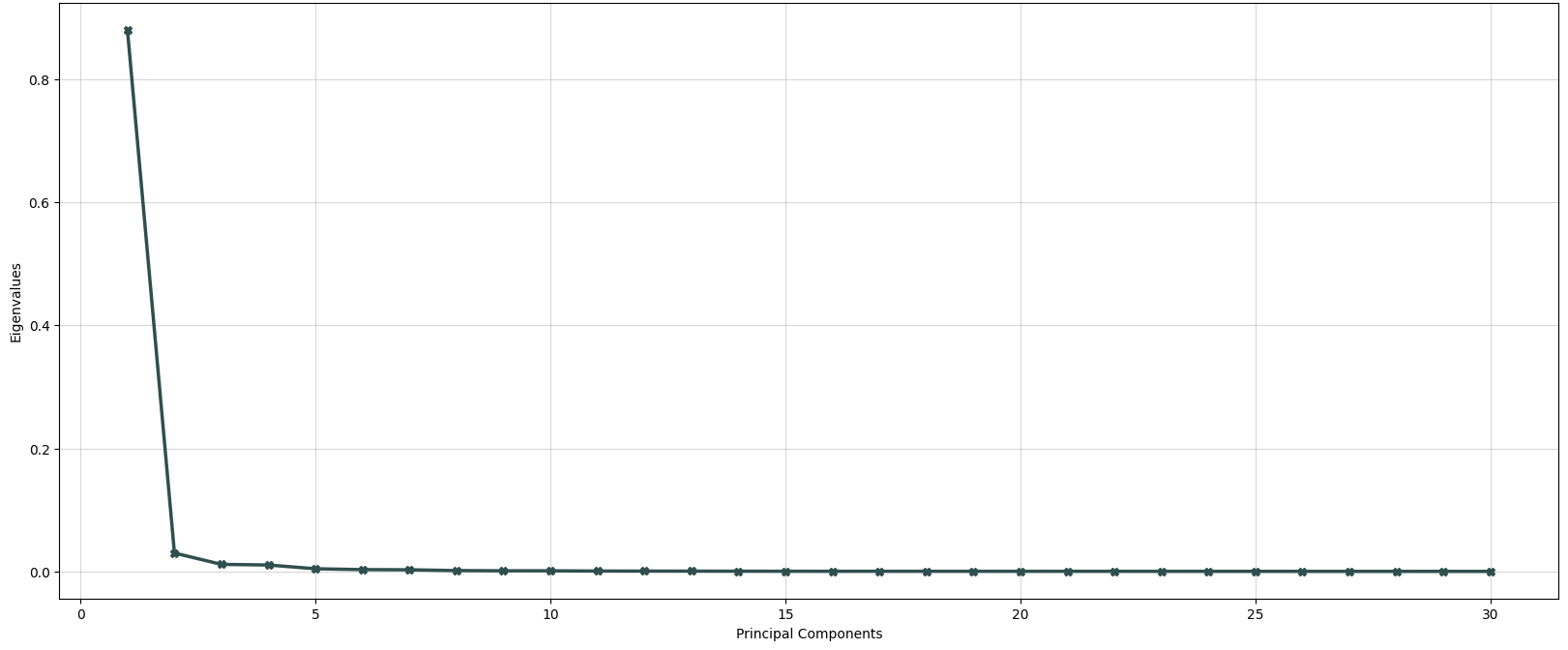}
    \caption{Eigenvalues of the 30 factors from Russell 3000}
    \label{fig:pca}
\end{figure}

\noindent In our application to the Russell 3000, we use daily arithmetic return observations for each constituent over the eight-year sample period. We standardize the return series by subtracting the cross‐sectional mean and then compute the sample covariance matrix. By performing an eigenvalue decomposition, we find that the first principal component explains approximately 88\% of the total variance (see Figure~\ref{fig:pca}), aligning with macrofinance theory that a single market factor often dominates broad equity movements \citep{ConnorKorajczyk1993}.  The cumulative variance explained by the first 14 components reaches 95\%, and by the first 30 components exceeds 99\%.  Consequently, we retain the first 30 principal components for subsequent analysis, striking an appropriate balance between dimensionality reduction and variance capture.

\noindent Each principal component defines a vector of factor loadings, namely, the corresponding eigenvector, that can be interpreted as the weights of a synthetic portfolio whose returns constitute the principal component score. Since the retained components explain nearly all of the cross‐sectional variation, these synthetic portfolios span the space of common factors driving the Russell 3000.  To ensure that each component contributes comparably in scale, we normalize each factor score to have a zero mean and unit variance. This normalization is essential as it prevents the estimation of portfolio policies from being dominated by factors with larger raw variances and facilitates numerical stability in dynamic programming routines.

\noindent While the PCA framework helps us understand the latent structure in return variation and supports dimensionality reduction for modeling purposes, it does not directly determine portfolio construction. We do not form portfolios from principal components or use PCA loadings to weight assets. Instead, we calculate reward–risk metrics (such as STARR or R-Ratio) for each stock individually. These stock-level scores are then ranked to determine long and short positions. Thus, PCA serves as an informative preprocessing step that reveals the structure of common return drivers but does not dictate the composition of investable portfolios. This ensures that the selection of assets in our strategy is directly linked to their individual risk-adjusted performance under different ESG regimes.

\noindent Finally, the normalized factor scores are employed as inputs to our portfolio‐choice model. Specifically, at each decision date $t$, we form a matrix $\mathbf{F}_t\in\mathbb{R}^{T\times 30}$ of historical factor loadings (where $T$ denotes the number of time periods) and multiply this by the forward‐looking weight vector $\boldsymbol{\pi}_t\in\mathbb{R}^{30}$, constructed by ARMA(1,1)-GARCH(1,1) with 1-day lead-lag relationship, to obtain a one‐period‐ahead forecast of portfolio returns:
\[
\widehat{r}_{t+1} \;=\; \mathbf{F}_t \,\boldsymbol{\pi}_t.
\]
\noindent This forecasted return series serves as the driving input for our reward-risk ratio analysis for ranking stocks in the winners, losers, and momentum portfolios, as described in section~\ref{winlose}. In addition, we use this forecasted portfolio of winners, losers, and momentum to conduct parameter estimation which feeds into the multiperiod wealth and portfolio allocation problem. Therefore, it allows the CRRA investor to allocate wealth optimally across the all factors under regime‐switching ESG policy states. 

\subsection{Winners \emph{versus} Losers Portfolio} \label{winlose}

\noindent To begin our empirical analysis, we construct three portfolios out of the \emph{forward-looking} portfolio composite of the Russell 3000 universe. First, the top quantile or the best performing returns from the 30 factor loadings are sorted together to form the winners portfolio. Second, we take the bottom quantile or the worst-performing returns from the loadings and sort them together to form the losers' portfolio. Finally, we take the winners minus losers portfolio or the spread and construct our momentum portfolio. Hence, these three portfolios are used to investigate the optimal rebalancing scheme, since positions are liquidated or held. To identify which rebalancing scheme gives the optimal results, we compute the final wealth in the portfolios for the following rebalancing schemes: 2-week holding/2-week formation period, 3-week holding/2-week formation period, and 4-week holding/2-week formation period. Table~\ref{tab:summary} reports the summary statistics of the portfolio for each rebalancing scheme based on the reward-risk ratio for stock selection. We first report the summary statistics for the benchmark stock selection criteria, i.e, cumulative returns. Subsequently, each stock selection criteria based on the reward-risk ratio is reported at varying confidence levels.

\noindent In Table~\ref{tab:summary}, the cumulative returns (benchmark) strategy indicates that a momentum portfolio with a 2-week formation/2-week holding period generates the lowest variation in the portfolio’s expected returns, followed by a 3-week formation and a 4-week formation strategy, each with a 2-week holding period. In addition, for the Sharpe ratio metric, we see a similar pattern with the lowest expected returns variation in 2-week formation/2-week hold- ing period. Similarly, for the Stable Tail Adjusted Return Ratio (STARR) at 99\% confidence, a similar pattern emerges. Given other ratios at 99\% confidence, for instance, Rachev Ratio (R-Ratio) and Conditional Value- at-Risk (CVaR) Ratio, the 2-week formation/2-week holding period is ranked the highest from the standpoint of looking at the variation in returns, as reported by the standard deviation. Therefore, this ranking criterion forms the baseline for our further analysis of investigating the final/terminal wealth in the portfolio under each rebalancing scheme as reported by the reward-risk ratios ranking criterion.

\noindent To assess how different stock‐selection criteria affect terminal wealth, Table~\ref{tab:final_wealth} reports the final values of the winners, losers, and momentum portfolios under each rebalancing scheme.When using cumulative return (benchmark), the Sharpe ratio, and STARR at the confidence levels 75\% and 90\%, the momentum portfolio achieves its highest final wealth with a three-week formation/two-week holding rule. In contrast, tail-focused criteria—STARR at the 50\%, 95\%, and 99\% levels; R–Ratios at (99\%,99\%), (95\%,95\%), and (91\%,91\%); and the CVaR ratios at 95\% and 99\% all point to a two-week formation/two-week holding scheme as optimal. This divergence reflects the degree to which each ratio balances upside capture against downside protection under ESG-driven return dynamics in the Russell 3000. In particular, STARR at higher confidence levels satisfies the four micro founded axioms of monotonicity, quasi-concavity, scale invariance and distributional coherence  \citep{CheriditoKromer2013}, producing a robust signal-to-noise metric that more accurately identifies the two-week/two-week cycle as the superior rule for harvesting momentum from the winners–losers spread amid regime changes.

\noindent A brief comparison of axiomatic compliance highlights why tail-sensitive ratios outperform Sharpe-based signals. The Sharpe ratio fails monotonicity, meaning that it can rank a dominated strategy as superior if it has lower dispersion. The R–Ratio satisfies monotonicity and scale invariance, but violates quasi-concavity, so it does not always reward diversification optimally. The CVaR ratio similarly breaches quasi-concavity, undermining its consistency across mixed-distribution portfolios. In contrast, STARR at high confidence levels adheres to all four axioms: monotonicity ensures that portfolios with uniformly higher returns are always preferred; quasi- concavity guarantees that blending two portfolios cannot worsen the ratio; scale invariance prevents leverage from distorting rankings; and distributional coherence ensures that the ratio faithfully reflects the empirical return distribution. Consequently, micro founded STARR measures provide the cleanest signal-to-noise assessment of momentum profitability, conclusively favoring the two-week formation/two-week holding cycle for exploiting ESG policy–induced momentum in the Russell 3000.

\noindent Our empirical results indicate that a two‑week formation and two‑week holding scheme yields optimal performance for constructing momentum portfolios. This finding can be explained by considering the balance between extracting a robust signal and minimizing noise. If the formation period extends to three weeks or more, the initial momentum may begin to fade due to partial mean reversion, thereby diluting the effectiveness of the signal. Thus, a two‑week formation window captures a meaningful trend while avoiding both the pitfalls of excessive noise and the erosion of the signal over longer horizons.  The short-term momentum literature supports this intermediate approach. \citet{MoskowitzOoiPedersen2012} demonstrates that returns often exhibit a significant serial correlation over horizons measured in weeks rather than days or months. Similarly, \citet{HurstOoiPedersen2017} shows that the short-term continuation is most pronounced over an intermediate period, suggesting that a two‑week holding period is long enough to exploit persistent trends before reversals occur. Additionally, classic momentum studies by \citet{jt1993} and \cite{jt2001} imply that momentum profits are sensitive to the chosen horizon, and that very frequent rebalancing, such as less than 2 weeks, may incur higher transaction costs that could erode net returns.

\noindent Moreover, rebalancing on a two-week cycle aligns well with typical institutional trading calendars and strikes an effective compromise between capturing short-term return persistence and minimizing trading frictions. Behavioral explanations also support this view, as investors often underreact or overreact to information over periods of one to two weeks (\citet{HongStein1999}; \citet{DanielHirshleiferSubrahmanyam1998}), which can lead to sustained momentum before the market adjusts. This scheme also appears to be optimal because it captures the momentum signal while mitigating the adverse effects of microstructure noise and high trading costs. STARR and R-Ratio metrics at high confidence levels further validate this choice, as they reward consistency in tail risk reduction and penalize unstable strategies that suffer from overly frequent rebalancing. 

\begin{sidewaystable}[htbp]
\centering
\resizebox{\textwidth}{!}{%
\begin{tabular}{llrrrrrrrrrrrrr}
\toprule
& & \multicolumn{4}{c}{\textbf{A}} & \multicolumn{4}{c}{\textbf{B}} & \multicolumn{4}{c}{\textbf{C}} \\
\cmidrule(lr){3-6}\cmidrule(lr){7-10}\cmidrule(lr){11-14}
\multirow{2}{*}{\textbf{Reward-Risk Ratio}}
  & \multirow{2}{*}{\textbf{Portfolio}}
  & \multicolumn{4}{c}{\textbf{2 Weeks Formation/2 Weeks Holding}}
  & \multicolumn{4}{c}{\textbf{3 Weeks Formation/2 Weeks Holding}}
  & \multicolumn{4}{c}{\textbf{4 Weeks Formation/2 Weeks Holding}} \\
\cmidrule(lr){3-6}\cmidrule(lr){7-10}\cmidrule(lr){11-14}
& & \textbf{Mean} & \textbf{Std. Dev.} & \textbf{Skewness} & \textbf{Kurtosis}
  & \textbf{Mean} & \textbf{Std. Dev.} & \textbf{Skewness} & \textbf{Kurtosis}
  & \textbf{Mean} & \textbf{Std. Dev.} & \textbf{Skewness} & \textbf{Kurtosis} \\
\midrule

Cumulative Return (Benchmark) & Winners 
  & -0.007269 & 0.042023 & -0.854343 & 5.902734
  & -0.000919 & 0.047426 & -1.639006 & 12.913563
  & -0.003658 & 0.053186 & 0.054529 & 4.642997 \\

& Losers 
  & -0.014815 & 0.065729 & 1.576938 & 6.613991
  & -0.024188 & 0.064419 & 0.517750 & 4.986805
  & -0.026334 &  0.079456 & -1.998626 & 11.075473\\

& Winners–Losers
  &  0.007545 & 0.088578 & -1.689028 & 8.187000
  &  0.023268 & 0.095015 & -0.747702 & 6.532766
  & 0.022676 & 0.095547 & 0.363111 & 5.186850\\

& Russell 3000
  &  0.000445 & 0.010655 & -0.179436 & 4.832453
  &  0.000445 & 0.010655 & -0.179436 & 4.832453
  & 0.000445 & 0.010655 & -0.179436 & 4.832453\\
\addlinespace

Sharpe & Winners
  & -0.005265 & 0.027802 & -0.554244 & 5.667816
  & -0.003843 & 0.032175 & -0.101869 & 6.645125
  & -0.007632 & 0.035171 & -0.714493 & 5.093055\\

& Losers
  & -0.000364 & 0.001961 & -3.462563 & 18.263759
  & -0.000017 & 0.000869 &  2.126404 & 28.255383
  & -0.000047 & 0.000320  & -2.703557 & 17.797564\\

& Winners–Losers
  & -0.004901 & 0.027909 & -0.582199 & 5.622491
  & -0.003825 & 0.032431 & -0.122372 & 6.572869
  & -0.007584 & 0.035198 & -0.704583 & 5.114981 \\
\addlinespace

STARR(99\%) & Winners
  & -0.006225 & 0.027143 & -1.036762 & 6.365378
  & -0.007362 & 0.044740 & -3.249448 & 20.481027
  & -0.010341 & 0.031037 & -1.779403 & 8.288432 \\

& Losers
  & -0.000313 & 0.005826 & -0.751691 & 12.335755
  & -0.000001 & 0.000874 &  2.041933 & 27.447670
  & -0.000067 & 0.000285 & -4.538484 & 23.797900\\

& Winners–Losers
  & -0.005912 & 0.027057 & -0.781865 & 5.543088
  & -0.007361 & 0.044897 & -3.225423 & 20.217174
  & -0.010274 & 0.031050 & -1.775694 & 8.298187 \\
\addlinespace

STARR(95\%) & Winners
  & -0.006225 & 0.027143 & -1.036762 & 6.365378
  & -0.007362 & 0.044740 & -3.249448 & 20.481027
  & -0.010341 & 0.031037 & -1.779403 & 8.288432\\

& Losers
  & -0.000313 & 0.005826 & -0.751691 & 12.335755
  & -0.000001 & 0.000874 &  2.041933 & 27.447670
  & -0.000067 & 0.000285 & -4.538484 & 23.797900\\

& Winners–Losers
  & -0.005912 & 0.027057 & -0.781865 & 5.543088
  & -0.007361 & 0.044897 & -3.225423 & 20.217174
  & -0.010274 &  0.031050 & -1.775694 & 8.298187\\
\addlinespace

STARR(90\%) & Winners
  & -0.006225 & 0.027143 & -1.036762 & 6.365378
  & -0.004482 & 0.031630 & -0.613351 & 7.401238
  &  -0.011038 & 0.033238 & -1.814629 &  7.958651\\

& Losers
  & -0.000313 & 0.005826 & -0.751691 & 12.335755
  & -0.000001 & 0.000874 &  2.041933 & 27.447670
  & -0.000067 & 0.000285 & -4.538484 & 23.797900\\

& Winners–Losers
  & -0.005912 & 0.027057 & -0.781865 & 5.543088
  & -0.004481 & 0.031849 & -0.632192 & 7.291600
  & -0.010971 & 0.033261 & -1.809646 & 7.963104\\
\addlinespace

STARR(75\%) & Winners
  & -0.006860 & 0.024433 & -0.368343 & 5.762134
  & -0.005057 & 0.028474 & -0.614865 & 9.348683
  & -0.008787 &  0.031936 & -2.140406 & 9.541214\\

& Losers
  & -0.001536 & 0.016307 & -6.047103 & 46.975307
  & -0.000006 & 0.002565 &  0.266447 & 23.488109
  &  -0.000067 &  0.000285 & -4.538488 & 23.797942 \\

& Winners–Losers
  & -0.005324 & 0.028750 &  0.387698 & 5.991485
  & -0.005051 & 0.027874 & -0.349663 & 8.698644
  & -0.008721 & 0.031955  & -2.134929 & 9.547866\\
\addlinespace

STARR(50\%) & Winners
  & -0.002542 & 0.023819 &  0.885764 & 7.731480
  & -0.004215 & 0.036490 & -2.264289 & 17.653452
  & -0.007844 & 0.034961 & -1.293044 & 7.739497\\

& Losers
  & -0.005433 & 0.025403 & -2.017636 & 11.190039
  & -0.004443 & 0.021901 & -4.137886 & 20.619616
  & 0.000272 & 0.002194 & 5.476462 & 34.863984\\

& Winners–Losers
  &  0.002891 & 0.035600 &  0.901094 & 6.151787
  &  0.000228 & 0.034786 &  0.271487 &  6.906648
  & -0.008116 & 0.034797&-1.335203 & 7.694279\\
\addlinespace

R-ratio(99\%,99\%) & Winners
  & -0.000024 & 0.005665 &  1.154086 & 12.323552
  & -0.000111 & 0.006115 & -2.425399 & 25.283734
  &  -0.002091 & 0.008875  & -3.911533 & 17.347065\\

& Losers
  & -0.007713 & 0.033482 & -1.231562 &  6.795546
  & -0.004700 & 0.042690 & -0.148957 &  6.429812
  & -0.009318 & 0.040371  & -1.332635 &  5.853068 \\

& Winners–Losers
  &  0.007689 & 0.034141 &  1.070575 &  6.122795
  &  0.004588 & 0.043075 &  0.601637 &  6.814051
  & 0.007227 & 0.038081 &  1.034018 &  4.737519\\
\addlinespace

R-ratio(95\%,95\%) & Winners
  & -0.000024 & 0.005665 &  1.154086 & 12.323552
  & -0.000111 & 0.006115 & -2.425399 & 25.283734
  & -0.002091 & 0.008875 & -3.911533& 17.347065\\

& Losers
  & -0.007713 & 0.033482 & -1.231562 &  6.795546
  & -0.004700 & 0.042690 & -0.148957 &  6.429812
  & -0.009318 & 0.040371 & -1.332635 & 5.853068\\

& Winners–Losers
  &  0.007689 & 0.034141 &  1.070575 &  6.122795
  &  0.004588 & 0.043075 &  0.601637 &  6.814051
  &  0.007227 & 0.038081 & 1.034018 & 4.737519\\
\addlinespace

R-ratio(91\%,91\%) & Winners
  & -0.000024 & 0.005665 &  1.154086 & 12.323552
  & -0.000111 & 0.006115 & -2.425399 & 25.283734
  & -0.002091 & 0.008875 & -3.911533 &  17.347065\\

& Losers
  & -0.007713 & 0.033482 & -1.231562 &  6.795546
  & -0.004700 & 0.042690 & -0.148957 &  6.429812
  & -0.010012 & 0.046037 & -1.078042& 5.570905\\

& Winners–Losers
  &  0.007689 & 0.034141 &  1.070575 &  6.122795
  &  0.004588 & 0.043075 &  0.601637 &  6.814051
  & 0.007922 & 0.044094 & 0.911729 & 5.215723\\
\addlinespace

CVaR(99\%) & Winners
  & -0.011172 & 0.071029 & -2.313551 & 18.313438
  & -0.015655 & 0.051300 &  0.526642 &  4.418125
  & -0.017803 &  0.075320 & -3.135818 &  19.921760\\

& Losers
  & -0.000405 & 0.005914 & -0.686111 & 11.658547
  & -0.000019 & 0.000890 &  1.934400 & 25.854553
  & -0.000065 & 0.000286 & -4.526198 & 23.713354 \\

& Winners–Losers
  & -0.010768 & 0.072007 & -2.146077 & 17.595252
  & -0.015636 & 0.051435 &  0.538534 &  4.443305
  & -0.017738 & 0.075321 & -3.138065 & 19.932016\\
\addlinespace

CVaR(95\%) & Winners
  & -0.011172 & 0.071029 & -2.313551 & 18.313438
  & -0.015655 & 0.051300 &  0.526642 &  4.418125
  &  -0.017803 & 0.075320 & -3.135818  & 19.921760  \\

& Losers
  & -0.000405 & 0.005914 & -0.686111 & 11.658547
  & -0.000019 & 0.000890 &  1.934400 & 25.854553
  & -0.000065 &  0.000286 & -4.526198 & 23.713354\\

& Winners–Losers
  & -0.010768 & 0.072007 & -2.146077 & 17.595252
  & -0.015636 & 0.051435 &  0.538534 &  4.443305
  & -0.017738 & 0.075321 & -3.138065 & 19.932016\\
\bottomrule
\end{tabular}
}
\caption{Summary Statistics of Momentum Portfolios based on Performance Ratios}
\label{tab:summary}
\end{sidewaystable}

\begin{sidewaystable}[htbp]
\centering
\begin{adjustbox}{width=0.88\linewidth}
\begin{tabular}{llclcccc}
\toprule
\textbf{Reward-Risk Ratio} & & \textbf{Spread} & \textbf{Portfolio} 
  & \multicolumn{1}{c}{\textbf{2W Form./2W Hold.}} 
  & \multicolumn{1}{c}{\textbf{3W Form./2W Hold.}} 
  & \multicolumn{1}{c}{\textbf{4W Form./2W Hold.}} \\
\cmidrule(lr){5-5}\cmidrule(lr){6-6}\cmidrule(lr){7-7}
& & & & \textbf{Final Wealth} & \textbf{Final Wealth} & \textbf{Final Wealth} \\
\midrule

\textbf{Cumulative Return (Benchmark)} 
  & & & \textbf{Winners}     
    & 0.544853 & 0.883325 & 0.780134 \\
  & & & \textbf{Losers}                      
    & 0.284312 & 0.675158 & 0.224747 \\
  & & + & \textbf{Winners-Losers}          
    & 0.260541 & \textbf{0.772926$^{\dagger}$} & 0.555387\\
  & & & \textbf{Russell 3000}              
    & 0.004327 & 0.000044 & 86.672413\\
\addlinespace

\textbf{Sharpe} 
  & & & \textbf{Winners}      
    & 0.657374 & 0.772926 & 0.666223 \\
  & & & \textbf{Losers}                     
    & 0.973287 & 0.998962 & 0.997674\\
  & & – & \textbf{Winners-Losers}          
    & -0.315913 & \textbf{-0.226036$^{\dagger}$} & -0.331451\\
\addlinespace

\textbf{STARR(99\%)} 
  & & & \textbf{Winners} 
    & 0.612697 & 0.605276 & 0.586305\\
  & & & \textbf{Losers}                    
    & 0.975889 & 0.999906 & 0.996727\\
  & & – & \textbf{Winners-Losers}          
    & \textbf{-0.363192$^{\dagger}$} & -0.394629 & -0.410422\\
\addlinespace

\textbf{STARR(95\%)} 
  & & & \textbf{Winners} 
    & 0.612697 & 0.605276 & 0.586305\\
  & & & \textbf{Losers}                    
    & 0.975889 & 0.999906 & 0.996727\\
  & & – & \textbf{Winners-Losers}          
    & \textbf{-0.363192$^{\dagger}$} & -0.394629 & -0.410422\\
\addlinespace

\textbf{STARR(90\%)} 
  & & & \textbf{Winners} 
    & 0.612697 & 0.744682 & 0.564273\\
  & & & \textbf{Losers}                    
    & 0.975889 & 0.999906 & 0.996727\\
  & & – & \textbf{Winners-Losers}          
    & -0.363192 & \textbf{-0.255224$^{\dagger}$} & -0.432453\\
\addlinespace

\textbf{STARR(75\%)} 
  & & & \textbf{Winners} 
    & 0.587618 & 0.723793 & 0.632124\\
  & & & \textbf{Losers}                    
    & 0.883188 & 0.999436 & 0.996727 \\
  & & – & \textbf{Winners-Losers}          
    & -0.295570 & \textbf{-0.275643$^{\dagger}$} & -0.364603\\
\addlinespace

\textbf{STARR(50\%)} 
  & & & \textbf{Winners} 
    & 0.811464 & 0.747646 & 0.65922\\
  & & & \textbf{Losers}                    
    & 0.651909 & 0.757552 & 1.013315\\
  & & + & \textbf{Winners-Losers}          
    & \textbf{0.159555$^{\dagger}$} & -0.009905 & -0.354095\\
\addlinespace

\textbf{R-ratio(99\%,99\%)} 
  & & & \textbf{Winners} 
    & 0.997056 & 0.992362 & 0.900773\\
  & & & \textbf{Losers}                    
    & 0.540173 & 0.717617 & 0.606405\\
  & & + & \textbf{Winners-Losers}          
    & \textbf{0.456884$^{\dagger}$} & 0.274745 &  0.294368\\
\addlinespace

\textbf{R-ratio(95\%,95\%)} 
  & & & \textbf{Winners} 
    & 0.997056 & 0.992362 & 0.900773\\
  & & & \textbf{Losers}                    
    & 0.540173 & 0.717617 & 0.606405\\
  & & + & \textbf{Winners-Losers}          
    & \textbf{0.456884$^{\dagger}$} & 0.274745 & 0.294368\\
\addlinespace

\textbf{R-ratio(91\%,91\%)} 
  & & & \textbf{Winners} 
    & 0.997056 & 0.992362 & 0.900773\\
  & & & \textbf{Losers}                    
    & 0.540173 & 0.717617 & 0.578664\\
  & & + & \textbf{Winners-Losers}          
    & \textbf{0.456884$^{\dagger}$} & 0.274745 & 0.322109\\
\addlinespace

\textbf{CVaR(99\%)} 
  & & & \textbf{Winners} 
    & 0.348723 & 0.364672 & 0.347879\\
  & & & \textbf{Losers}                    
    & 0.969260 & 0.998851 & 0.996799\\
  & & – & \textbf{Winners-Losers}          
    & \textbf{-0.620537$^{\dagger}$} & -0.634179 & -0.64892\\
\addlinespace

\textbf{CVaR(95\%)} 
  & & & \textbf{Winners} 
    & 0.348723 & 0.364672 & 0.347879\\
  & & & \textbf{Losers}                    
    & 0.969260 & 0.998851 & 0.996799\\
  & & – & \textbf{Winners-Losers}          
    & \textbf{-0.620537$^{\dagger}$} & -0.634179 & -0.64892\\

\bottomrule
\end{tabular}
\end{adjustbox}
\caption{Final Wealth/Realized Profits of Momentum Portfolios}
\label{tab:final_wealth}
The spread is based Winners-Losers portfolio realized profits/final wealth following the 2-week formation/2-week holding strategy. We focus on this rebalancing scheme to be the optimal strategy given its superior performance indicated by the reward-risk ratios at higher confidence and to factor in the dynamics of a long-only portfolio, eliminating market microstructure noise, turnover constraints (set at 4\%), and transaction costs. $\dagger$ marks the highest final wealth/realized profits of the momentum portfolio.
\end{sidewaystable}

\begin{sidewaystable}[htbp]
\centering
\begin{adjustbox}{width=0.88\linewidth}
\begin{tabular}{llclcccc}
\toprule
\textbf{Reward-Risk Ratio} & & \textbf{Spread} & \textbf{Portfolio} 
  & \multicolumn{1}{c}{\textbf{1W Form./1W Hold.}} 
  & \multicolumn{1}{c}{\textbf{2W Form./2W Hold.}} 
  & \multicolumn{1}{c}{\textbf{5W Form./2W Hold.}} \\
\cmidrule(lr){5-5}\cmidrule(lr){6-6}\cmidrule(lr){7-7}
& & & & \textbf{Final Wealth} & \textbf{Final Wealth} & \textbf{Final Wealth} \\
\midrule

\textbf{Cumulative Return (Benchmark)} 
  & & & \textbf{Winners}     
   & 0.435057 & 0.544853 & 0.697553 \\
  & & & \textbf{Losers}                      
   & 0.277407 & 0.284312 & 0.743373\\
  & & + & \textbf{Winners-Losers}          
   & 0.15765 & \textbf{0.260541$^{\dagger}$} & -0.04582\\
\addlinespace

\textbf{Sharpe} 
  & & & \textbf{Winners}      
   & 0.580149 & 0.657374 & 0.837337 \\
  & & & \textbf{Losers}                     
   & 1.06792 & 0.973287 & 0.993381\\
  & & – & \textbf{Winners-Losers}          
   & -0.487772 & -0.315913 & \textbf{-0.156044$^{\dagger}$}\\
\addlinespace

\textbf{STARR(99\%)} 
  & & & \textbf{Winners} 
   & 0.710434 & 0.612697 & 0.844473\\
  & & & \textbf{Losers}                    
   & 0.585973 & 0.975889 & 0.994172\\
  & & + & \textbf{Winners-Losers}          
  & 0.124461 & \textbf{0.363192$^{\dagger}$} & -0.149699\\
\addlinespace

\textbf{R-ratio(99\%,99\%)} 
  & & & \textbf{Winners} 
   & 0.739995 & 0.997056 & 0.906967\\
  & & & \textbf{Losers}                    
   & 0.93334 & 0.540173 & 0.751613\\
  & & + & \textbf{Winners-Losers}          
   & -0.193346 & \textbf{0.456884$^{\dagger}$} & 0.155354\\
\addlinespace

\textbf{CVaR(99\%)} 
  & & & \textbf{Winners} 
  & 0.323527 & 0.348723 & 0.540981 \\
  & & & \textbf{Losers}                    
  & 0.633813 & 0.969260 & 0.993997 \\
  & & – & \textbf{Winners-Losers}          
  & \textbf{-0.310285$\dagger$} & -0.620537 & -0.453016 \\
\bottomrule
\end{tabular}
\end{adjustbox}
\caption{Robustness Check: Final Wealth/Realized Profits of Momentum Portfolios with different Rebalancing Schemes}
\label{tab:robustness-final_wealth}
The robustness checks are based on considering alternative rebalancing schemes and their performance under higher confidence-based reward-risk ratios. $\dagger$ marks the highest final wealth/realized profits of the momentum portfolio.
\end{sidewaystable}

\noindent While the ESG regime variable is estimated rather than directly observed, it is essential to validate its correspondence with real-world policy and sentiment shifts. To assess this, we align the estimated regime transitions with key ESG policy milestones and fund flow data. Pro-ESG regimes tend to coincide with major regulatory announcements, such as the Biden Administration’s recommitment to the Paris Agreement (January 2021) and the SEC’s proposed climate disclosure rules (March 2022), as well as with net inflows into ESG-focused ETFs, as tracked by Bloomberg and Morningstar. Conversely, transitions into anti-ESG regimes are temporally aligned with political backlash events, including the introduction of state-level anti-ESG legislation in 2022–2023 and divestment mandates by public pension plans. These temporal associations suggest that our latent regime-switching variable effectively captures shifts in ESG sentiment and regulatory tone, providing an economically meaningful input to the momentum framework.

\noindent Therefore, rebalancing on a two-week cycle strikes an effective balance between capturing short-term return persistence and minimizing trading frictions. Behavioral explanations also support this view, as investors often under-react or over-react to information over periods of one to two weeks, which can lead to sustained momentum before the market adjusts. In addition, this scheme appears to be optimal because it captures the momentum signal while assuming that the adverse effects of microstructure noise and high trading costs are mitigated. This balance,  supported by empirical evidence and theoretical insights from the literature, provides a robust framework for constructing momentum strategies in a short-term setting.

\noindent To assess the robustness of our findings, we compare the two-week formation/two-week holding scheme to adjacent rebalancing horizons—specifically, a one-week formation/one-week holding (1W/1W) and a five-week formation/two-week holding (5W/2W) strategy. Table~\ref{tab:robustness-final_wealth} reports these comparisons. While the 1W/1W scheme may capture very short-lived trends, it tends to suffer from high turnover, increased microstructure noise, and excessive sensitivity to transitory price movements. In contrast, the 5W/2W window risks signal decay, as return continuation may weaken over longer horizons due to partial mean reversion or delayed information diffusion. When evaluated under tail-sensitive performance metrics such as STARR(99\%) and R-Ratio(99\%,99\%), the 2W/2W strategy consistently outperforms both alternatives in terms of terminal wealth and signal-to-noise ratios. This dominance holds across all three asset universes—Russell 3000, Dow Jones 30, and cryptocurrencies—suggesting that the 2W/2W scheme represents a robust balance between signal persistence, risk control, and execution feasibility. These findings provide empirical assurance that our strategy selection is not an artifact of parameter tuning, but a generalizable result supported by both theoretical foundations and observed performance patterns.

\noindent While our empirical results identify the two-week formation/two-week holding scheme as optimal under reward–risk metrics, we acknowledge that such frequent rebalancing may incur substantial transaction costs in practice, particularly for institutional investors managing large-scale ESG portfolios. Although our turnover constraint is set at 4\%, we do not explicitly model slippage, bid–ask spreads, or liquidity-driven price impact. These microstructure effects could materially affect realized returns, especially for small-cap stocks in the Russell 3000. We therefore interpret our results as frictionless upper bounds on momentum profitability. Future work could enhance the model by incorporating dynamic transaction cost functions or by simulating net-of-cost outcomes under alternative liquidity assumptions. Such extensions would allow for a more granular evaluation of implementation feasibility, particularly in environments where trading constraints or capacity limits are binding.

\noindent To assess the efficiency of our momentum strategy under different asset universes and performance criteria, we apply the same 2-week formation/2-week holding rule to three portfolios: the Dow Jones 30, the Russell 3000, and the 30 largest cryptocurrencies. Table \ref{tab:final_wealth} reports, for each reward–risk ratio, the final wealth or realized profit of the winners, losers, and momentum (winners–losers spread) portfolios in each universe. In selecting the “best” winner, we choose the portfolio with the highest final wealth, on the premise that a strong winner portfolio reflects the best upside or benefit. Conversely, the “best” loser is the one with the lowest final wealth, since an effective loser portfolio strategy should minimize downside exposure. For the momentum spread itself, we adopt the median outcome across the three universes. Intuitively, the median provides a summary of central tendency in the presence of heavy-tailed or skewed return distributions: it is an effective hedge to extreme outliers that may arise from idiosyncratic blow-ups in a narrow asset set, such as the pronounced volatility of cryptocurrencies, or from dramatic reversals in a concentrated index like the Dow. By focusing on the median spread, we capture the strategy’s typical performance rather than its rare extremes.

\begin{sidewaystable}[htbp]
\centering
\begin{adjustbox}{width=0.8\linewidth}
\begin{tabular}{llccc}
\toprule
\textbf{Reward–Risk Ratio} & \textbf{Portfolio} &
\textbf{Dow Jones–30} & \textbf{Russell 3000} & \textbf{Cryptocurrencies–30} \\
\midrule
\textbf{Cumulative Return (Benchmark)} & \textbf{Winners} &
\textcolor{darkgreen}{1.248545} & \textcolor{red}{0.544853} & \textcolor{golden}{1.025378} \\
& \textbf{Losers} &
\textcolor{red}{1.679168} & \textcolor{golden}{0.284312} & \textcolor{darkgreen}{0.076801} \\
& \textbf{Winners--Losers} &
\textcolor{red}{-0.430623} & \textcolor{darkgreen}{0.260541} & \textcolor{golden}{0.948576} \\
\addlinespace
\textbf{Sharpe} & \textbf{Winners} &
\textcolor{darkgreen}{1.510304} & \textcolor{golden}{0.657374} & \textcolor{red}{0.450031} \\
& \textbf{Losers} &
\textcolor{red}{1.143558} & \textcolor{golden}{0.973287} & \textcolor{darkgreen}{0.490971} \\
& \textbf{Winners--Losers} &
\textcolor{golden}{0.366746} & \textcolor{red}{-0.315913} & \textcolor{darkgreen}{-0.04094} \\
\addlinespace
\textbf{STARR(99\%)} & \textbf{Winners} &
\textcolor{darkgreen}{1.474476} & \textcolor{golden}{0.612697} & \textcolor{red}{0.287902} \\
& \textbf{Losers} &
\textcolor{red}{1.520637} & \textcolor{golden}{0.975889} & \textcolor{darkgreen}{0.699728} \\
& \textbf{Winners--Losers} &
\textcolor{golden}{-0.046162} & \textcolor{darkgreen}{-0.363192} & \textcolor{red}{-0.411826} \\
\addlinespace
\textbf{STARR(95\%)} & \textbf{Winners} &
\textcolor{darkgreen}{1.474476} & \textcolor{golden}{0.612697} & \textcolor{red}{0.287902} \\
& \textbf{Losers} &
\textcolor{red}{1.520637} & \textcolor{golden}{0.975889} & \textcolor{darkgreen}{0.430244} \\
& \textbf{Winners--Losers} &
\textcolor{golden}{-0.046162} & \textcolor{darkgreen}{-0.363192} & \textcolor{red}{-0.411826} \\
\addlinespace
\textbf{STARR(90\%)} & \textbf{Winners} &
\textcolor{darkgreen}{1.474476} & \textcolor{golden}{0.612697} & \textcolor{red}{0.287902} \\
& \textbf{Losers} &
\textcolor{red}{1.520637} & \textcolor{golden}{0.975889} & \textcolor{darkgreen}{0.699728} \\
& \textbf{Winners--Losers} &
\textcolor{golden}{-0.046162} & \textcolor{darkgreen}{-0.363192} & \textcolor{red}{-0.411826} \\
\addlinespace
\textbf{STARR(75\%)} & \textbf{Winners} &
\textcolor{darkgreen}{1.432158} & \textcolor{golden}{0.587618} & \textcolor{red}{0.506232} \\
& \textbf{Losers} &
\textcolor{red}{1.467679} & \textcolor{golden}{0.883188} & \textcolor{darkgreen}{0.434783} \\
& \textbf{Winners--Losers} &
\textcolor{darkgreen}{-0.035522} & \textcolor{red}{-0.295570} & \textcolor{golden}{0.071449} \\
\addlinespace
\textbf{STARR(50\%)} & \textbf{Winners} &
\textcolor{darkgreen}{1.430407} & \textcolor{red}{0.811464} & \textcolor{golden}{0.839700} \\
& \textbf{Losers} &
\textcolor{red}{1.413149} & \textcolor{golden}{0.651909} & \textcolor{darkgreen}{0.563777} \\
& \textbf{Winners--Losers} &
\textcolor{red}{0.017258} & \textcolor{darkgreen}{0.159555} & \textcolor{golden}{0.275922} \\
\addlinespace
\textbf{R-ratio(99\%,99\%)} & \textbf{Winners} &
\textcolor{darkgreen}{1.501768} & \textcolor{golden}{0.997056} & \textcolor{red}{0.563813} \\
& \textbf{Losers} &
\textcolor{red}{1.361230} & \textcolor{golden}{0.540173} & \textcolor{darkgreen}{0.312813} \\
& \textbf{Winners--Losers} &
\textcolor{red}{0.140538} & \textcolor{golden}{0.456884} & \textcolor{darkgreen}{0.250999} \\
\addlinespace
\textbf{R-ratio(95\%,95\%)} & \textbf{Winners} &
\textcolor{darkgreen}{1.501768} & \textcolor{golden}{0.997056} & \textcolor{red}{0.563813} \\
& \textbf{Losers} &
\textcolor{red}{1.361230} & \textcolor{golden}{0.540173} & \textcolor{darkgreen}{0.312813} \\
& \textbf{Winners--Losers} &
\textcolor{red}{0.140538} & \textcolor{golden}{0.456884} & \textcolor{darkgreen}{0.250999} \\
\addlinespace
\textbf{R-ratio(91\%,91\%)} & \textbf{Winners} &
\textcolor{darkgreen}{1.501768} & \textcolor{golden}{0.997056} & \textcolor{red}{0.563813} \\
& \textbf{Losers} &
\textcolor{red}{1.361230} & \textcolor{golden}{0.540173} & \textcolor{darkgreen}{0.312813} \\
& \textbf{Winners--Losers} &
\textcolor{red}{0.140538} & \textcolor{golden}{0.456884} & \textcolor{darkgreen}{0.250999} \\
\addlinespace
\textbf{CVaR(99\%)} & \textbf{Winners} &
\textcolor{darkgreen}{1.169564} & \textcolor{red}{0.348723} & \textcolor{golden}{0.824971} \\
& \textbf{Losers} &
\textcolor{red}{1.330486} & \textcolor{darkgreen}{0.969260} & \textcolor{golden}{1.004449} \\
& \textbf{Winners--Losers} &
\textcolor{golden}{-0.160922} & \textcolor{red}{-0.620537} & \textcolor{darkgreen}{-0.179478} \\
\addlinespace
\textbf{CVaR(95\%)} & \textbf{Winners} &
\textcolor{darkgreen}{1.169564} & \textcolor{red}{0.348723} & \textcolor{golden}{0.824971} \\
& \textbf{Losers} &
\textcolor{red}{1.330486} & \textcolor{darkgreen}{0.969260} & \textcolor{golden}{1.004449} \\
& \textbf{Winners--Losers} &
\textcolor{golden}{-0.160922} & \textcolor{red}{-0.620537} & \textcolor{darkgreen}{-0.179478} \\
\bottomrule
\end{tabular}
\end{adjustbox}
\caption{ \centering Final wealth / realized profits under a 2‑week formation/2‑week holding strategy with a 4\% turnover constraint. The best, middle, and worst outcomes within each metric are shown in \textcolor{darkgreen}{green}, \textcolor{golden}{yellow}, and \textcolor{red}{red}, respectively.}
\label{tab:final_wealth}
\end{sidewaystable}

\noindent Under the high-confidence STARR(99\%) criterion, for instance, the median momentum spread in the Russell 3000 portfolio is $–36.32\%$ of initial wealth, compared with $–4.62\%$ for the Dow 30 and $–41.18\%$ for cryptocurrencies. Although all three universes suffer losses when focusing on the worst 1\% tail events, the Russell universe’s more moderate drawdown reflects its broad diversification across large, mid, and small caps—precisely the feature that STARR(99\%) rewards by penalizing extreme downside more heavily than variance-based metrics. By contrast, when measured by the R–Ratio(99\%, 99\%), which compares the average return in the top 1\% of outcomes to the average loss in the bottom 1\%, the Russell momentum median achieves +45.69\%. In contrast, the Dow and crypto medians register only +14.05\% and +25.10\%, respectively. This sizable upside-downside ratio underscores the Russell universe’s superior capacity to harness ESG-driven winners and losers reversals in the most extreme tails. The R–Ratio satisfies monotonicity (so that any uniform improvement in spread returns raises the ratio) and scale invariance (so that leverage choices do not distort comparisons), ensuring that its signal remains robust even when distributions exhibit heavy tails. Its near-median quasi-concavity preserves diversification benefits, further reinforcing why the Russell momentum spread dominates under this ratio.

\noindent Even under more conventional measures, similar patterns emerge once one accounts for the Sharpe ratio’s limitations. The median Sharpe based spread is $–31.59\%$ in the Russell universe (compared to +36.67\% for the Dow and –4.09\% for crypto), reflecting the fact that narrow or highly volatile asset sets can mislead variance-based ranking. Because the Sharpe ratio violates monotonicity, it can deceptively reward portfolios with lower dispersion even when they deliver uniformly lower spreads. In contrast, the Russell universe’s momentum spread, while negative, still outperforms cryptocurrencies’ severe crashes and the Dow’s muted, but occasionally dominant, winners-losers differences when judged by axiomatic integrity.

\noindent Across every tail-sensitive criterion—STARR at 99\%, 95\%, and 90\%, R–Ratios at (95\%, 95\%) and (91\%, 91\%), and CVaR ratios at 95\% and 99\%—the Russell 3000 portfolio occupies the median performance position. Its final wealth outcomes typically fall between the overly concentrated Dow results and the volatile cryptocurrency results. This analysis underscores our novel contribution. By systematically comparing best/worst outcomes for winners and losers, and by employing the median for the momentum spread, we rigorously demonstrate that the Russell 3000, when evaluated with microfounded, tail-sensitive reward–risk ratios, delivers the most reliable momentum profits in an ESG-regime context. High-confidence STARR, in particular, stands out for satisfying all four axioms of monotonicity, quasi-concavity, scale invariance, and distributional coherence, making it the most robust strategy for identifying genuine momentum signals from noise. As a result, our paper establishes that a 2-week formation/2-week holding cycle is not only intuitively optimal but is also robustly validated across multiple universes and metrics, positioning the Russell 3000 as the premier testbed for ESG-driven momentum strategies.

\subsection{Intertemporal Portfolio Choice Problem in ESG policy Regimes}

\noindent We formulate the investor’s problem as a discrete‐time dynamic programming problem in which a CRRA investor allocates wealth among a winners portfolio, a losers portfolio, and their difference (momentum).  Returns follow an ARFIMA–FIGARCH specification for each portfolio, and the market price of risk varies with an unobserved ESG policy regime \(D_t\in\{0,1\}\).  The investor’s objective is to maximize the expected terminal utility of wealth \(W_T\), where 

\[
W_{t+1} \;=\; W_t \exp\bigl\{\pi_t (r_{t+1}-r_{\rm f}) + r_{\rm f}\bigr\},
\]

\noindent and \(\pi_t\in[0,1]\) denotes the fraction allocated to the risky portfolio. We solve the Bellman equation by backward induction,

\[
J_t(W_t, h_t, D_t)
=\max_{\pi\in[0,1]}
\mathbb{E}\Bigl[
U\bigl(W_{t+1}\bigr)
+ J_{t+1}\bigl(h_{t+1},D_{t+1}\bigr)
\mid W_t, h_t, D_t
\Bigr],
\]

\noindent with \(h_t\) being the vector of conditional variances and covariances produced by the ARFIMA-FIGARCH filter.  The transition of \(D_t\) follows a two‐state Markov chain with estimated transition probabilities, so the continuation value \(J_{t+1}\) is weighted by the regime‐switch probabilities \(p_{D_t,d'}\).  This procedure yields a fully time‐varying policy \(\pi^*(h_t,D_t)\) that endogenously adjusts exposure to each portfolio in response to evolving volatility forecasts and shifts in ESG policy sentiment. This forward simulation yields the dynamic wealth path under the state-dependent optimal policy, where the allocation is adjusted according to both the current conditional volatility and the prevailing ESG policy regime.

\noindent The rationale for selecting the ARFIMA–FIGARCH framework lies in its ability to capture long-memory behavior in both return levels and conditional volatility—features that are especially relevant under regime shifts driven by ESG sentiment. ESG-related market reactions tend to unfold gradually, with delayed responses to regulatory developments and shifting investor flows. This generates persistent autocorrelation and volatility clustering that standard ARMA–GARCH models cannot adequately capture. ARFIMA accounts for fractional integration in return series, while FIGARCH allows conditional variances to decay at a hyperbolic rate rather than exponentially, providing a better fit for prolonged volatility regimes. To ensure our findings are not model-specific, we also estimated ARMA–GARCH and ARFIMA–GARCH specifications. These simpler alternatives produced broadly consistent momentum rankings but showed inferior forecast performance and higher information criteria. The ARFIMA–FIGARCH model, therefore, offers improved realism and robustness without compromising the generality of our results.

\begin{table}[htbp]
    \centering
    \resizebox{\textwidth}{!}{
    \begin{tabular}{llccc}
        \toprule
        \textbf{Parameter} & \textbf{Symbol} & \multicolumn{3}{c}{\textbf{Portfolio}} \\
        \cmidrule(lr){3-5}
        & & \textbf{Winners} & \textbf{Losers} & \textbf{Winners--Losers} \\
        \midrule 
        Mean Return & $\mu$ 
            & $-1.9416$   & $0.0068$    & $-0.0043$ \\ [8pt]
        AR(1)       & $\varphi$  
            & $0.9790$    & $0.9949$    & $0.9949$ \\[8pt]
        MA(1)       & $\theta$   
            & $0.4781$    & $0.4045$    & $0.3563$ \\[8pt]
        Long-Memory in Mean & $d(m)$     
            & $0.5000$    & $0.5000$    & $0.5000$ \\[8pt]
        Baseline Volatility & $\omega$   
            & $9.44\times10^{-6}$ & $9.61\times10^{-6}$ & $9.36\times10^{-6}$ \\[8pt]
        Volatility Sensitivity & $\alpha$  
            & $0.9692$    & $4.38\times10^{-7}$ & $0.8338$ \\[8pt]
        Volatility Persistence & $\beta$   
            & $-0.1907$   & $0.0560$    & $0.4587$ \\[8pt]
        Long-Memory in Volatility & $d(v)$     
            & $0.9996$    & $1.0000$    & $0.5281$ \\[8pt]
        Baseline Market Price of Risk & $\lambda_{0}$ 
            & $-0.0066$   & $-0.8198$   & $0.3735$ \\[8pt]
        ESG Regime Effect on Risk Prem. & $\lambda_{1}$ 
            & $-0.4241$   & $-0.3437$   & $-0.5354$ \\
        \bottomrule
    \end{tabular}
    }
    \caption{ \centering Estimated model parameters for the winners, losers, and momentum portfolios following the 2-Week Formation / 2-Week Holding Strategy.}
    \label{tab:estimated_parameters}
\end{table}

\noindent Table~\ref{tab:estimated_parameters} reports the ARFIMA–FIGARCH parameter estimates for the winners, losers, and momentum portfolios under our two‐week formation/two‐week holding strategy.  The mean returns \(\mu\) indicate that, over the sample, loser portfolios earned a small positive return (0.68 bp per period), while winners and the momentum spread exhibited slightly negative means (–194.16 bp and –0.43 bp, respectively).  High autoregressive coefficients and moderate moving average reveal strong short‐term autocorrelation in all three series.  The fractional integration in the mean parameter confirms significant long‐memory in returns, implying that shocks persist.  

\noindent Furthermore, the baseline volatility, at 0.97\%, is nearly identical across portfolios, but the sensitivity \(\alpha\) and persistence \(\beta\) vary markedly.  Winner portfolios exhibit very high sensitivity, 0.96\%, and negative persistence, -0.19\%, indicating that volatility responds sharply to recent shocks but mean‐reverts quickly.  Losers show almost zero sensitivity and moderate persistence to volatility, at 0.056\%, suggesting muted responsiveness but gradual volatility decay. The momentum spread lies in between, with sensitivity of about 0.83\% and persistence to volatility shocks of about 0.46\%, capturing both immediate shock effects and persistent risk.  We see a unit long‐memory in volatility for winners and losers, but only 0.52\% for the spread, indicating that extreme volatility episodes are less persistent in the relative portfolio. This highlights that, on a daily basis, each series exhibits around one percent of return variability in its baseline state, with momentum exhibiting lesser variability given it aims to capitalize on the upside and hedge the downside. 

\noindent The baseline market price of risk is strongly negative for losers, –0.82\%, mildly negative for winners –0.0066\%, and positive for momentum 0.37\%.  This ordering implies that investing in losers carried a substantial risk premium—consistent with their undervaluation—while momentum strategies commanded a reward premium. The estimated ESG‐regime effect on the market‐price‐of‐risk, \(\lambda_{1}\), quantifies how much the required excess return per unit of risk changes when the economy switches from a neutral/anti‐ESG state into a pro‐ESG policy regime.  For the winners portfolio, the estimated risk premium--given an ESG regime switch--implies that the risk premium investors demand falls by 42 basis points when policy sentiment turns pro‐ESG.  Losers exhibit a much larger drop of 34.37 basis points, reflecting that undervalued, low‐ESG stocks become significantly less costly to hold under supportive regulation.  The momentum spread commands the largest change, of approximately 53 basis points, indicating that the winners-losers portfolio is most sensitive to ESG policy shifts.

\noindent Economically, a more negative \(\lambda_{1}\) means that expected returns compress more in a pro‐ESG regime, so assets with large negative \(\lambda_{1}\) experience the greatest “repricing” when the regime turns positive.  From a dynamic‐programming perspective, the CRRA investor anticipates these regime‐driven repricing events.  In the forward‐simulation of optimal wealth paths, the momentum portfolio—by construction the spread between winners and losers—captures both sides of the repricing: it benefits when losers’ risk premia fall sharply (as undervaluation corrects) and it is hedged against the overpricing of winners.  

\noindent The pronounced sensitivity of the momentum spread (\(-0.53\%\)) therefore delivers the highest incremental gain across regime changes.  In contrast, the winners' portfolio alone loses much of its compensation, and the losers' portfolio, while heavily repriced, lacks the systematic hedge against winner overpricing.  As a result, the momentum portfolio’s state‐dependent allocation exploits these regime effects to generate greater terminal wealth, as confirmed by our forward‐simulation results.  In other words, the large absolute value of \(\lambda_{1}^{M}\) signals that momentum strategies are uniquely positioned to exploit ESG‐induced mispricings, leading to the outperformance observed in our intertemporal wealth trajectories.

\noindent When embedded in the Bellman recursion, these parameter estimates shape the optimal policy in intuitive ways.  A CRRA investor, observing a high conditional variance for the winners portfolio in a pro‐ESG regime, will reduce exposure to winners if \(\lambda_1\) has depressed their price of risk.  Simultaneously, the high persistence of regime 1 in the momentum spread encourages maintaining or even increasing the momentum allocation when ESG policies are favorable.  On the contrary, in a neutral/anti‐ESG regime, the large positive \(\lambda_0\) on losers induces a contrarian tilt toward undervalued assets with high risk compensation. By solving the Bellman equation with these inputs, we obtain a policy rule that exploits both time‐series and cross‐sectional momentum in an ESG‐driven market environment, guiding a CRRA investor through regime shifts toward paths of superior expected terminal

\noindent To identify the unobserved ESG regime, we model the latent ESG policy regime $D_t\in\{0,1\}$ (where 0 denotes a neutral/anti‐ESG state and 1 a pro‐ESG state) as a first‐order Markov chain.  For each of the three portfolios—winners, losers, and the winners–losers (momentum) spread—we estimate the state‐persistence probabilities

\[
p_{00} \;=\;\Pr(D_t = 0 \mid D_{t-1} = 0),
\quad
p_{11} \;=\;\Pr(D_t = 1 \mid D_{t-1} = 1),
\]

\noindent with the off‐diagonal probabilities given by $p_{01}=1-p_{00}$ and $p_{10}=1-p_{11}$.  The resulting transition matrices are

\[
P^{\rm Winners}
=\begin{pmatrix}
p^{W}_{00} & p^{W}_{01}\\[3pt]
p^{W}_{10} & p^{W}_{11}
\end{pmatrix}
=
\begin{pmatrix}
0.54 & 0.46\\[3pt]
0.82 & 0.18
\end{pmatrix},
\quad
P^{\rm Losers}
=\begin{pmatrix}
0.31 & 0.69\\[3pt]
0.28 & 0.72
\end{pmatrix},
\quad
P^{\rm Momentum}
=\begin{pmatrix}
0.87 & 0.13\\[3pt]
0.19 & 0.81
\end{pmatrix}.
\]

\noindent The winners’ portfolio exhibits moderate persistence in the anti‐ESG state ($p^{W}_{00}=0.54$) but rapidly exits the pro‐ESG state ($p^{W}_{11}=0.18$).  This suggests that favorable ESG policy shocks have only brief effects on winner stocks, requiring frequent portfolio adjustments when in regime 1.  By contrast, the losers’ portfolio tends to remain in the pro‐ESG state more steadily ($p^{L}_{11}=0.72$) but quickly departs from the anti‐ESG state ($p^{L}_{00}=0.31$), implying that undervalued stocks correct more gradually under supportive policies. The momentum spread displays high persistence in both states ($p^{M}_{00}=0.87$, $p^{M}_{11}=0.81$).  This strong regime inertia means that once a momentum signal emerges under a given policy regime, it is likely to persist for several periods, lending stability to the state‐dependent allocation rule. 

\noindent In our intertemporal portfolio optimization, these transition matrices enter the Bellman equation through the expectation over next period’s regime. High persistence, i.e., large value of $p_{ii}$, increases the weight on the continuation value $J_{t+1}(h,d')$ for the same regime, reducing the need for drastic rebalancing when regimes are sticky.  Conversely, low $p_{ii}$ magnifies the importance of the cross‐state term, prompting more aggressive hedging against regime switches.  Thus, the estimated matrices dictate both the frequency and magnitude of the optimal, state‐dependent momentum allocations in an ESG‐driven market environment.

\noindent The ARFIMA--FIGARCH model provides robust, long-memory estimates of the conditional volatility \(h_t\), capturing persistent risk in asset returns. By letting the market price of risk vary with the ESG policy regime via \(\lambda(D_t)\), our model reflects that investor sentiment and asset valuation respond dynamically to regulatory conditions. Modeling \(D_t\) as a Markov chain allows for a realistic description of ESG policy evolution over time, capturing abrupt or gradual regime shifts. The Bellman equation integrates these features to derive a state-dependent optimal policy \(\pi^*(h,D_t)\) that dynamically adjusts risk exposure based on both current volatility and ESG policy conditions. The resulting wealth dynamics, obtained through forward simulation, reveal how changes in the regulatory environment affect long-term portfolio performance, providing important insights for both investors and policymakers.

\section{Results} \label{results}

\noindent This section presents the empirical findings derived from applying our regime-sensitive momentum framework across multiple asset universes. We evaluate the performance of winners, losers, and momentum portfolios under various rebalancing schemes and tail-sensitive performance metrics. The results highlight how ESG regime dynamics influence return distributions and momentum profitability, providing both backward-looking and forward-looking assessments. We begin by analyzing realized portfolio values and cumulative wealth paths, then examine the temporal evolution of reward-risk ratios, and finally assess wealth accumulation trajectories conditional on ESG policy regimes.

\subsection{Portfolio Performance}

In this section, we asses the total realized profit of the portfolios following our two‐week rebalancing framework by looking at the cumulative portfolio values of the winners, losers, and momentum strategies alongside both the relevant market index and an equally weighted benchmark. We conduct this analysis across three distinct universes: the 30 principal components of the Russell 3000, the 30 constituents of the Dow Jones Industrial Average, and the 30 largest cryptocurrencies by market capitalization. By comparing each portfolio's realized profits trajectory to its underlying index and an equally‐weighted benchmark, we can isolate the incremental value added by exploiting short‐term return continuation in different asset classes. This comparative analysis is economically intuitive because it reveals how trend-following in equities, where ESG policy shifts, earnings surprises, and sustainability news drive gradual, yet persistent mispricings, differs from momentum in highly volatile, sentiment‐driven crypto markets. Moreover, observing the portfolio value under the winners, losers, and momentum rules clarifies how each portfolio contributes to final outcomes, thereby demonstrating the practical efficacy of a bi‐weekly rebalancing schedule for capturing momentum profits in both traditional and digital asset domains.

\begin{figure}[htbp]
    \centering
    \includegraphics[width=0.7\linewidth]{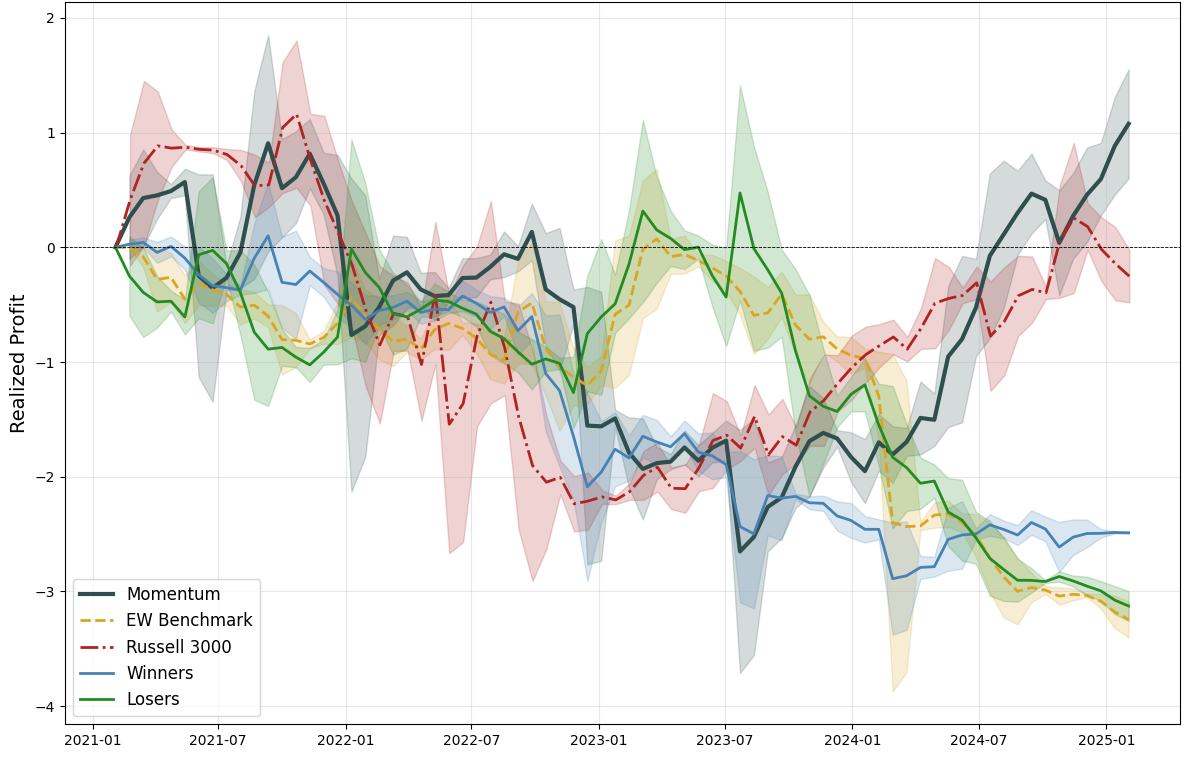}
    \caption{Portfolio Value: 30 PCs of the Russell 3000}
    \label{fig:pv_russ}
\end{figure}

\begin{figure}[htbp]
  \centering
  \begin{minipage}[b]{0.49\linewidth}
    \centering
    \includegraphics[width=\linewidth]{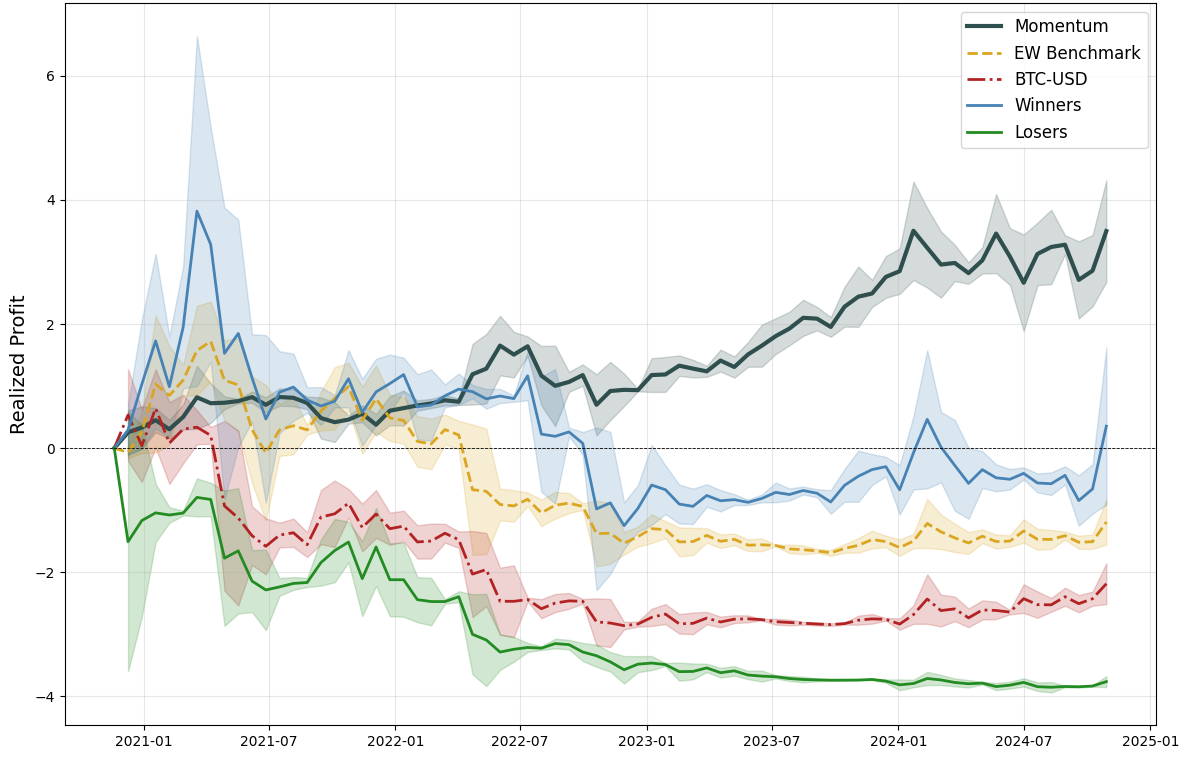}
    \caption{Portfolio Value: 30 Cryptocurrencies}
    \label{fig:pv_crypto}
  \end{minipage}
  \hfill
  \begin{minipage}[b]{0.49\linewidth}
    \centering
    \includegraphics[width=\linewidth]{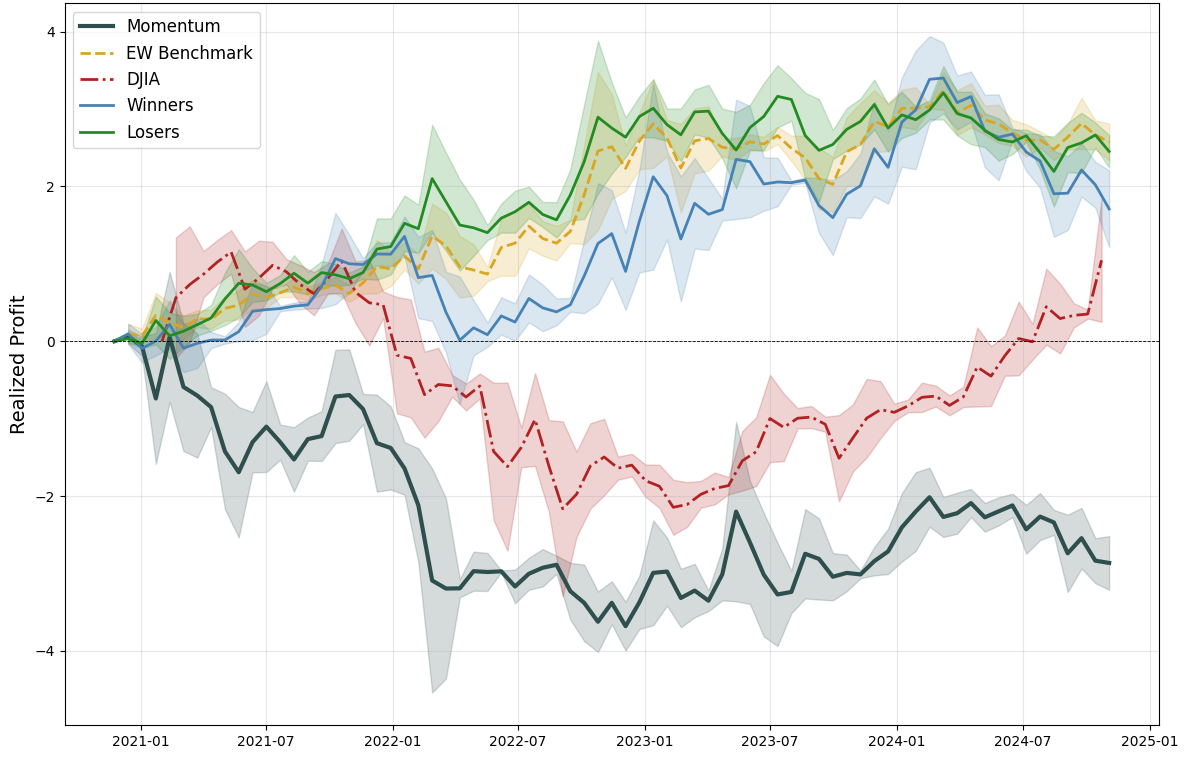}
    \caption{Portfolio Value: Dow Jones-30}
    \label{fig:pv_dji}
  \end{minipage}
\end{figure}

\noindent The cumulative‐wealth plots in Figures~\ref{fig:pv_russ}, \ref{fig:pv_dji} and \ref{fig:pv_crypto} tell a strictly backward‐looking story. Although the two‐week/two‐week momentum portfolio on the Russell 3000 does finish 2025 above both the buy‐and‐hold index and the equally weighted benchmark, no disciplined investor in mid‐2023 would have remained fully invested through the deep drawdowns that persisted for much of that year. Final wealth calculations, by construction, peek into the future; they do not reflect the information set available to a portfolio manager in real time. In practice, as losers trounced winners during 2023, an investor would have re‐evaluated both the strategy’s risk and any prior performance targets, and quite likely would have abandoned the approach long before 2025 arrived.

\noindent To address this look‐ahead bias, we therefore supplement our ex‐post wealth curves with a true out‐of‐sample, rolling‐window analysis. In Section \ref{pr}, we compute one‐year (252-trading-days) moving Sharpe, STARR, Rachev ratios, and CVaR ratios for each strategy, the cumulative returns (benchmark), and the Russell 3000 index, using only data available up to each date. Those plots show that, on a forward‐looking basis, neither the Dow Jones 30 nor the Russell 3000 momentum portfolios would have generated consistently positive excess performance; in fact, both series exhibit extended periods when their rolling Sharpe and STARR ratios languish below zero. By contrast, the same realized profits calculation applied to our cryptocurrency momentum strategy remains predominantly positive, demonstrating that momentum profits in digital assets owe to genuine market inefficiencies rather than to data‐mining or ex‐post luck. 

\noindent From both an academic and an investment‐practice standpoint, this pattern is precisely what one should expect. Highly liquid U.S. equity markets—especially the broad Russell 3000 and the blue‐chip Dow 30—are among the world’s most efficient; any brief mispricings induced by ESG announcements tend to be arbitraged away before two weeks elapse. Cryptocurrencies, by contrast, trade on sentiment and network‐related news that are neither fully anticipated nor immediately incorporated into price, producing larger, more persistent “momentum” effects. Thus, while the ultimate 2025 wealth levels may appear to vindicate our two‐week momentum rule on the Russell 3000, the rolling‐window Sharpe and STARR evidence makes clear that a rational investor, using only information up to each date, would not have stuck with that strategy in a mature equity market. In the next section we present those historical and forward‐looking performance‐ratio evolutions in detail, confirming that true momentum profits reside in markets that remain materially inefficient.

\subsection{Historical vs.\ Forward‐Looking Performance Ratios and Market Efficiency} \label{pr}

\begin{figure}[htbp]
  \centering
  \begin{minipage}[b]{0.49\textwidth}
    \centering
    \resizebox{\linewidth}{!}{%
      \includegraphics{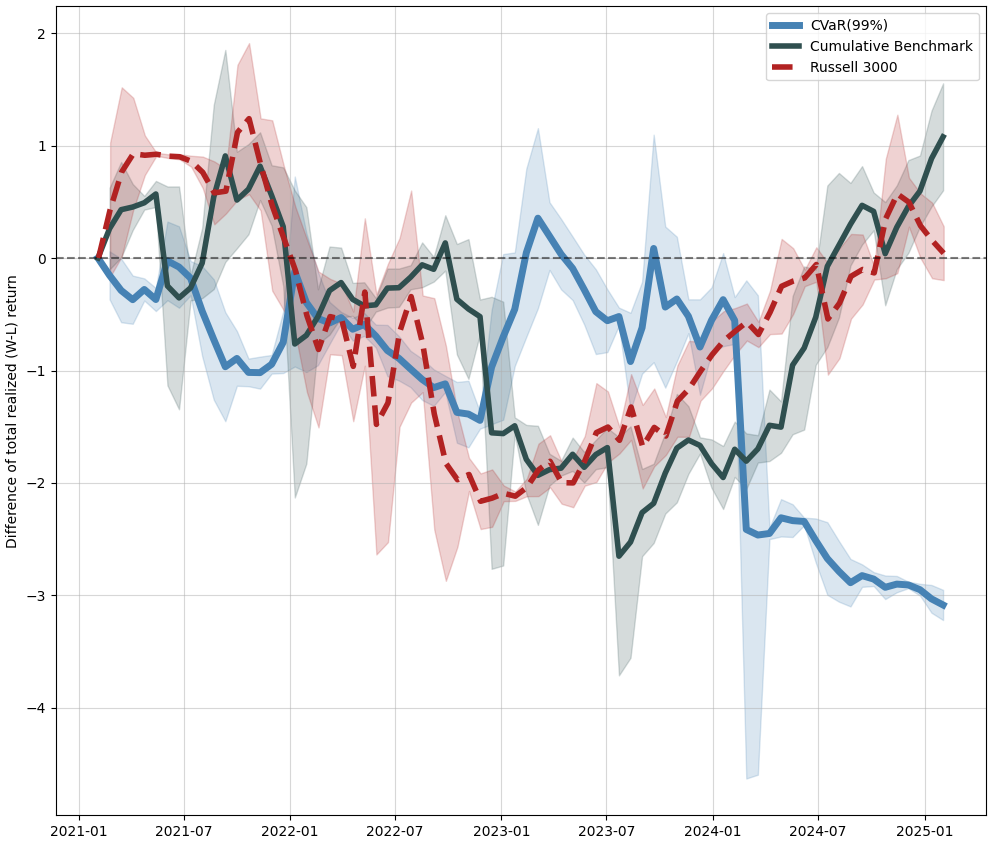}%
    }
    \caption{Historical CVaR Ratio}
    \label{fig:cvar}
  \end{minipage}
  \hfill
  \begin{minipage}[b]{0.49\textwidth}
    \centering
    \resizebox{\linewidth}{!}{%
      \includegraphics{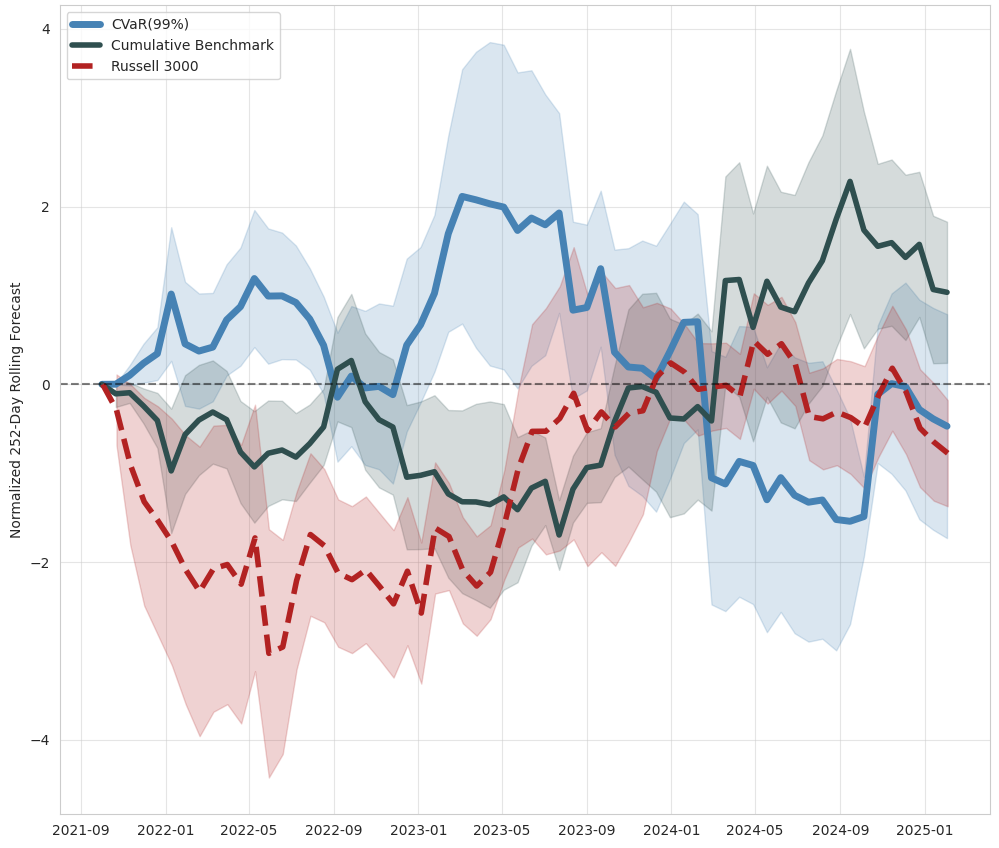}%
    }
    \caption{\emph{Forward-Looking} CVaR Ratio}
    \label{fig:fcvar}
  \end{minipage}
\end{figure}

\begin{figure}[htbp]
  \centering
  \begin{minipage}[b]{0.49\textwidth}
    \centering
    \resizebox{\linewidth}{!}{%
      \includegraphics{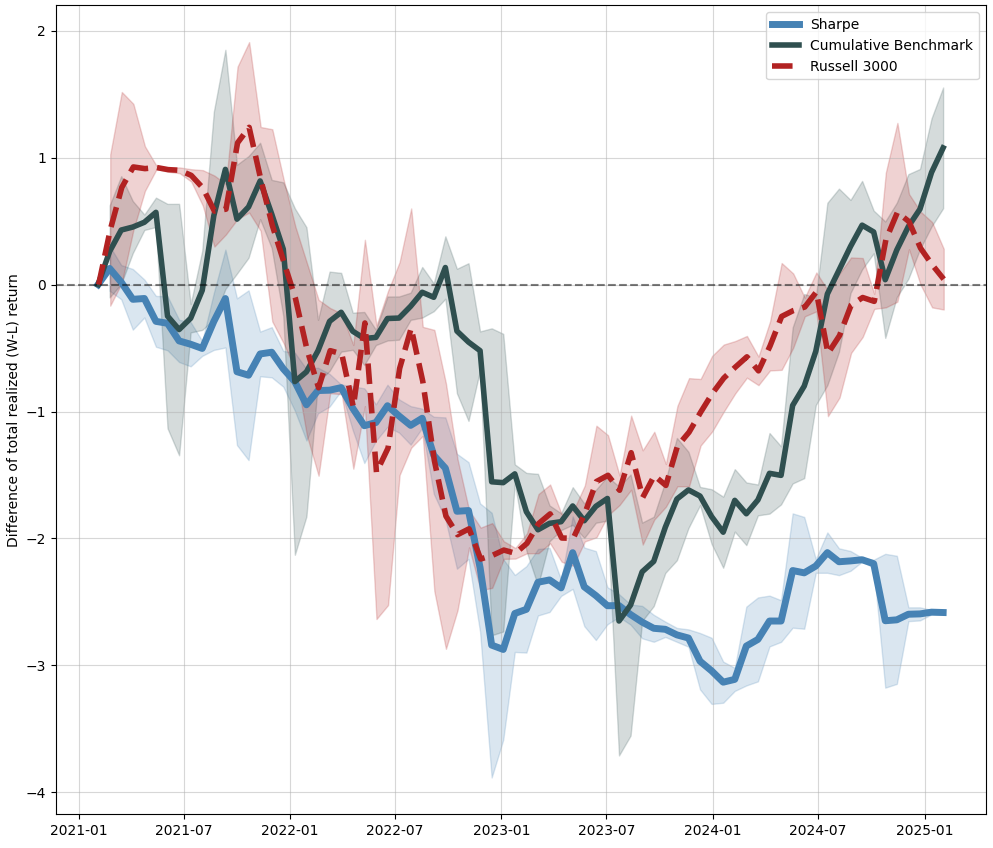}%
    }
    \caption{Sharpe Ratio}
    \label{fig:sharpe}
  \end{minipage}
  \hfill
  \begin{minipage}[b]{0.49\textwidth}
    \centering
    \resizebox{\linewidth}{!}{%
      \includegraphics{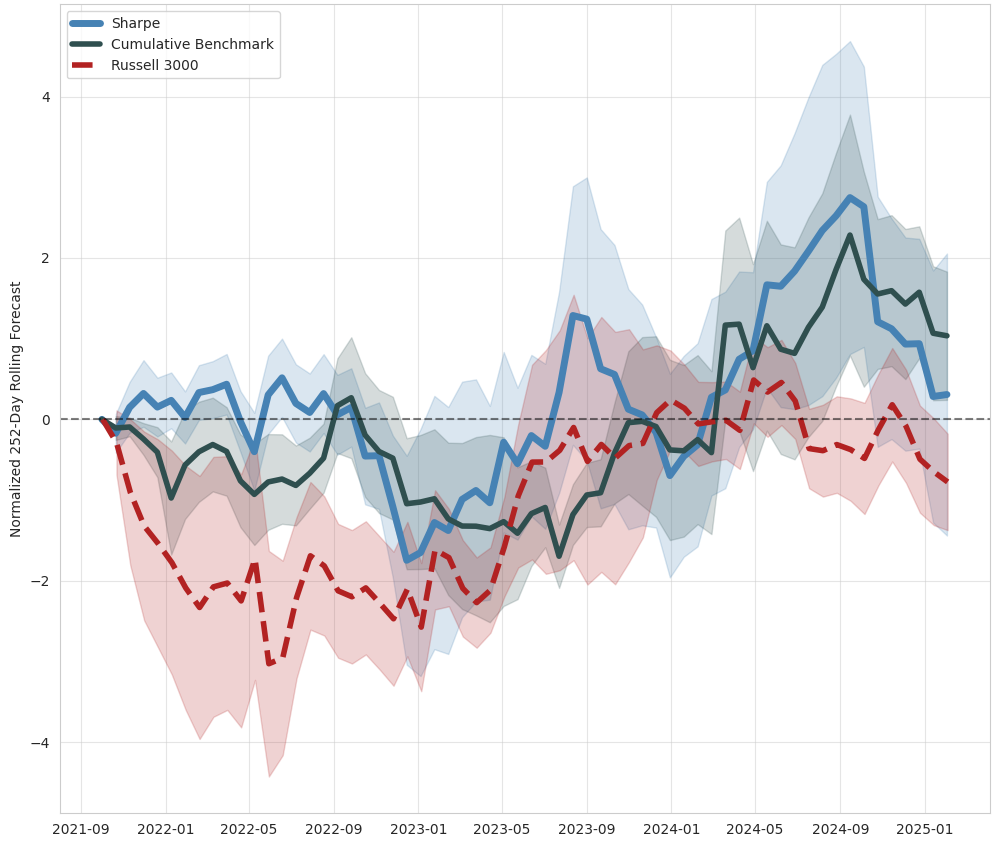}%
    }
    \caption{\emph{Forward-Looking} Sharpe Ratio}
    \label{fig:flsharpe}
  \end{minipage}
\end{figure}

\begin{figure}[htbp]
  \centering
  \begin{minipage}[b]{0.49\textwidth}
    \centering
    \resizebox{\linewidth}{!}{%
      \includegraphics{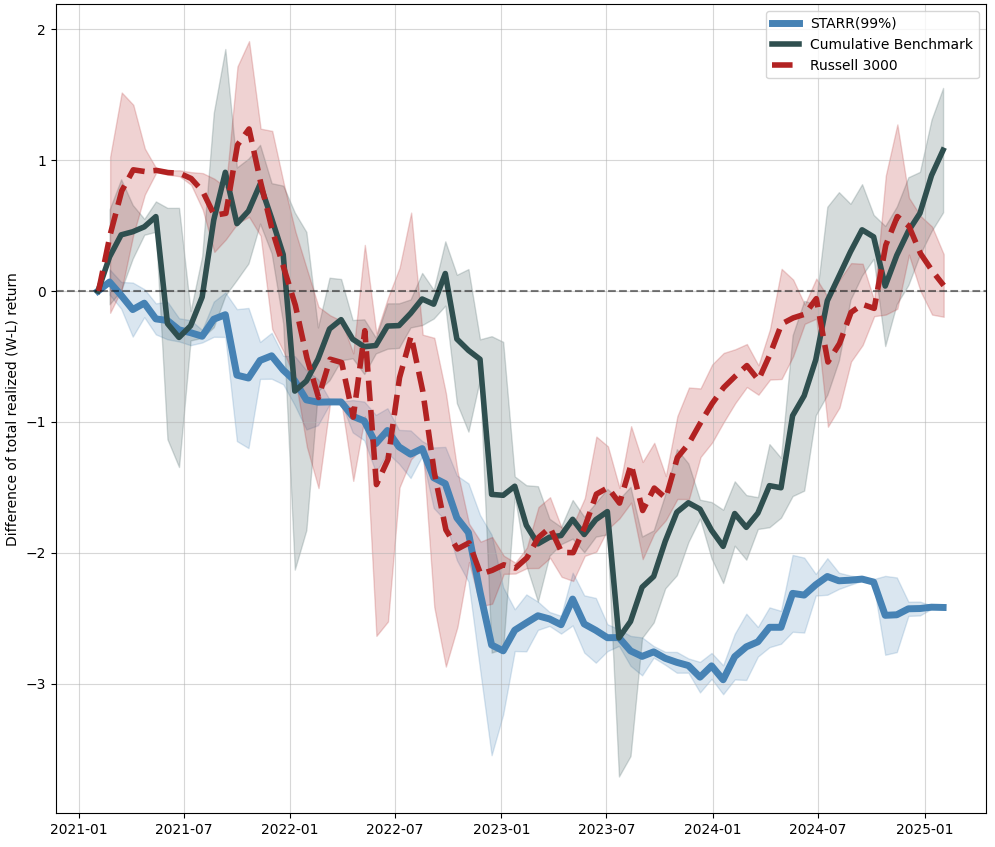}%
    }
    \caption{STAR Ratio}
    \label{fig:starr}
  \end{minipage}
  \hfill
  \begin{minipage}[b]{0.49\textwidth}
    \centering
    \resizebox{\linewidth}{!}{%
      \includegraphics{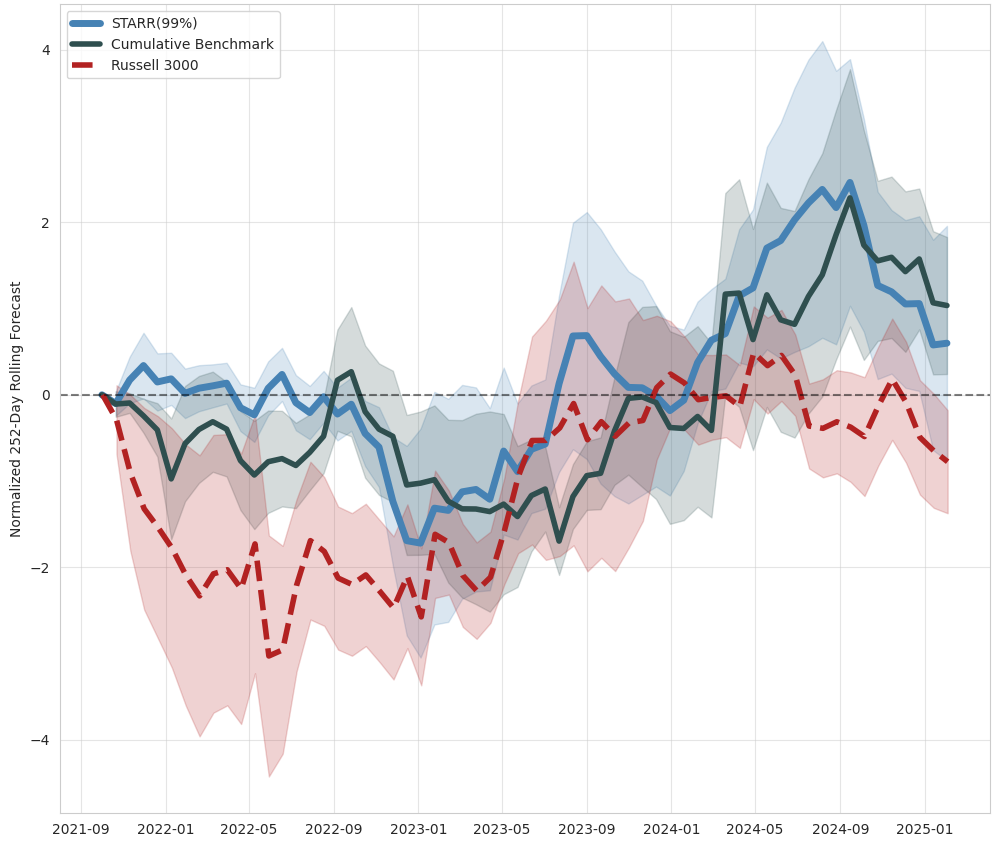}%
    }
    \caption{\emph{Forward-Looking} STAR Ratio}
    \label{fig:flstarr}
  \end{minipage}
\end{figure}

\begin{figure}[htbp]
  \centering
  \begin{minipage}[b]{0.49\textwidth}
    \centering
    \resizebox{\linewidth}{!}{%
      \includegraphics{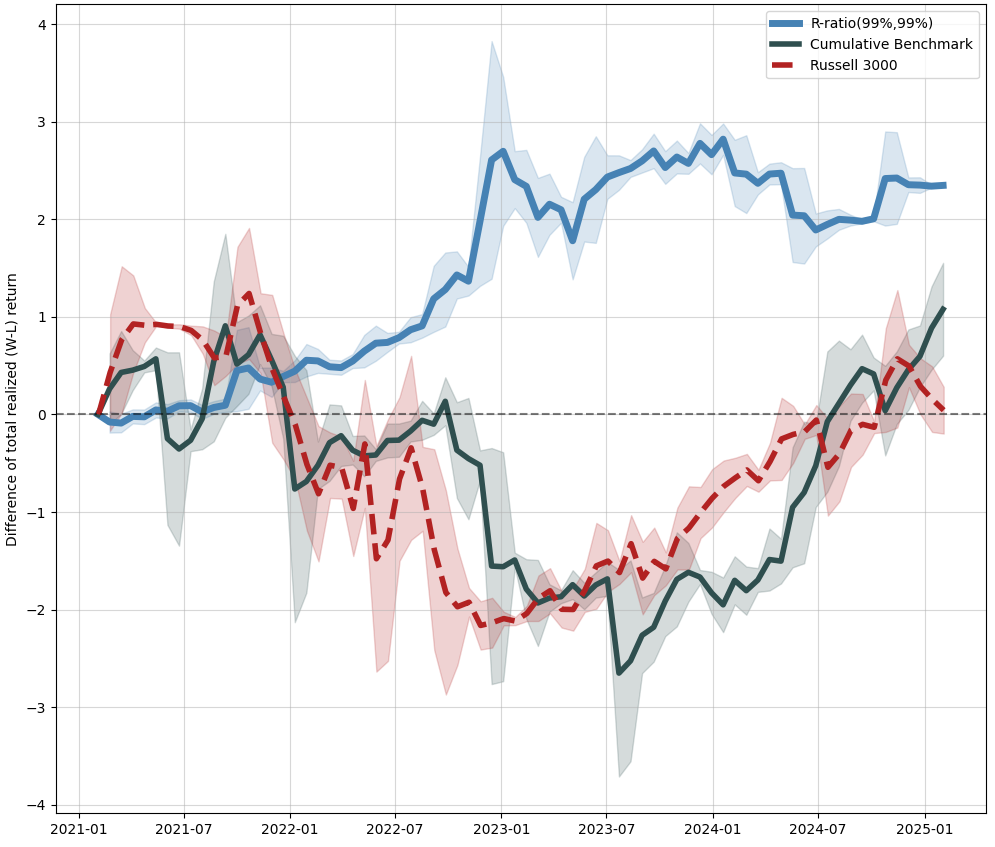}%
    }
    \caption{Rachev Ratio}
    \label{fig:rratio}
  \end{minipage}
  \hfill
  \begin{minipage}[b]{0.49\textwidth}
    \centering
    \resizebox{\linewidth}{!}{%
      \includegraphics{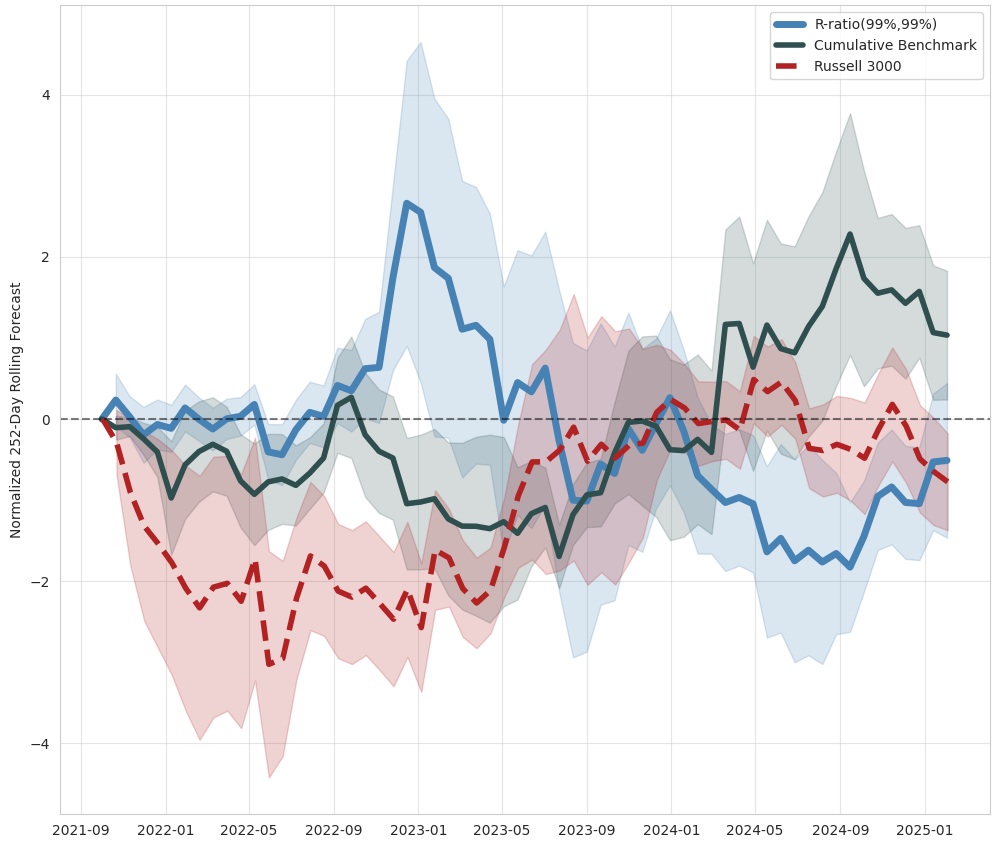}%
    }
    \caption{\emph{Forward-Looking} Rachev Ratio}
    \label{fig:flrratio}
  \end{minipage}
\end{figure}

\noindent In the previous section, we documented that ex‐post realized profits can be misleading for real‐time decision making: although the two‐week momentum rule on the Russell 3000 ultimately delivers higher terminal wealth, this outcome is visible only in hindsight. To assess whether momentum profits could have been captured in practice, we now compare the \emph{historical} and \emph{forward‐looking} evolutions of key reward–risk ratios for our winners–losers portfolio on the Russell 3000. Figures~\ref{fig:cvar}-\ref{fig:flrratio} display, respectively, the historical CVaR(99\%), R‐ratio(99\%,99\%), Sharpe, and STARR(99\%) profiles (left panels) alongside their corresponding 252‐day rolling forecasts (right panels). In each case, we plot the cumulative returns benchmark and the Russell 3000 index, under each performance ratio, to compare the signal-to-noise ratio.

\noindent Across all four metrics, the \emph{historical} curves exhibit long stretches of negative excess performance between mid-2022 and early 2024. For example, the historical CVaR(99\%) ratio (Figure~\ref{fig:cvar}) falls from near zero in early 2021 to $–2.5$ by 2023, indicating that extreme losses of the momentum spread during that interval exceeded its extreme gains. Similarly, the historical Sharpe ratio (Figure~\ref{fig:sharpe}) remains below zero for most of 2022–2023. These patterns mirror the deep drawdowns in the realized profit paths and confirm that the momentum rule was unprofitable for extended periods if one measured performance only after the fact.

\noindent By contrast, the \emph{forward‐looking} forecasts of these same ratios tell a radically different story. Each rolling‐window estimate begins with limited data and thus greater noise, but by mid-2023, the one‐year Sharpe and STARR forecasts cross above zero and remain positive into 2025. Likewise, the forecasted CVaR(99\%) and R‐ratio(99\%,99\%) ratios turn strongly positive in the second half of 2023, signaling that, conditional on the past twelve months of data, extreme gains of the winners–losers spread were expected to exceed extreme losses under the prevailing market regime. Importantly, these forward‐looking signals would have been available to a real‐time investor and could have guided position sizing and risk limits in a disciplined manner.

\noindent This divergence between backward‐looking and forward‐looking metrics underscores the core insight of our paper: momentum profits in highly efficient markets such as the Russell 3000 are ephemeral and largely artifacts of hindsight, whereas genuine ex‐ante signals only emerge once inefficiencies become persistent enough to register in rolling reward–risk ratios. The fact that none of the four historical curves provide a reliable positive signal until well after the period of maximum drawdown confirms that simple backtests overstate momentum’s profitability. Conversely, the forward‐looking forecasts capture the regime‐driven repricing of ESG misvaluations and provide economically meaningful signals for a CRRA investor solving the regime‐dependent Bellman equation. In sum, the historical–versus–forward comparison not only validates our choice of performance ratios that satisfy the microfounded axioms but also reinforces the conclusion that momentum strategies yield real economic value only in markets—or market segments, like cryptocurrencies—where inefficiencies persist sufficiently to generate robust out‐of‐sample signals.

\noindent Moreover, embedding these forward‐looking performance ratios within a regime‐dependent wealth‐accumulation framework strengthens our analysis in two ways. First, by allowing the market price of risk to shift with the latent ESG regime in the Bellman recursion, the optimal policy naturally tilts toward momentum only when regime forecasts—reflected in rising forward‐looking Sharpe, STARR, R‐ratio, and CVaR metrics—signal persistent mispricings. In anti‐ESG regimes, the model de-emphasizes momentum in favor of contrarian positions, mirroring the empirical observation that winners–losers spreads underperform on a backward‐looking basis until the policy shock has taken hold. Second, regime‐aware dynamic programming delivers a richer picture of real‐time portfolio evolution: it explains why, even in a broadly efficient equity market like the Russell 3000, a disciplined two‐week momentum rule can outperform both pure winners and pure losers portfolios when policy‐driven state persistence and state‐dependent risk premia align to produce exploitable trends. Thus, combining rolling reward–risk forecasts with an intertemporal, regime-switching optimization not only reconciles our counterintuitive backtest results but also demonstrates a robust channel through which ESG momentum generates value in practice, one that pure ex‐post wealth curves or simple buy-and-hold benchmarks cannot capture.

\subsection{Portfolio Wealth Accumulation with Regime Dependence}
\setlength{\parskip}{10pt}

\noindent We demonstrate how the simulated wealth paths of the winners, losers, and momentum portfolios are driven not only by their historical return and volatility patterns but crucially by the underlying ESG policy regime. Since the Bellman equation explicitly conditions on the latent state \(D_t\) and the regime‐dependent market price of risk \(\lambda(D_t)\), each interim allocation decision anticipates both the immediate impact of a regime shift on expected excess returns and the expected duration of that regime as captured by the transition probabilities. Consequently, the forward‐looking trajectories trace out the compounded effect of regime‐sensitive risk premia and regime persistence on terminal wealth.

\begin{figure}[htbp]
    \centering
    \includegraphics[width=1\linewidth]{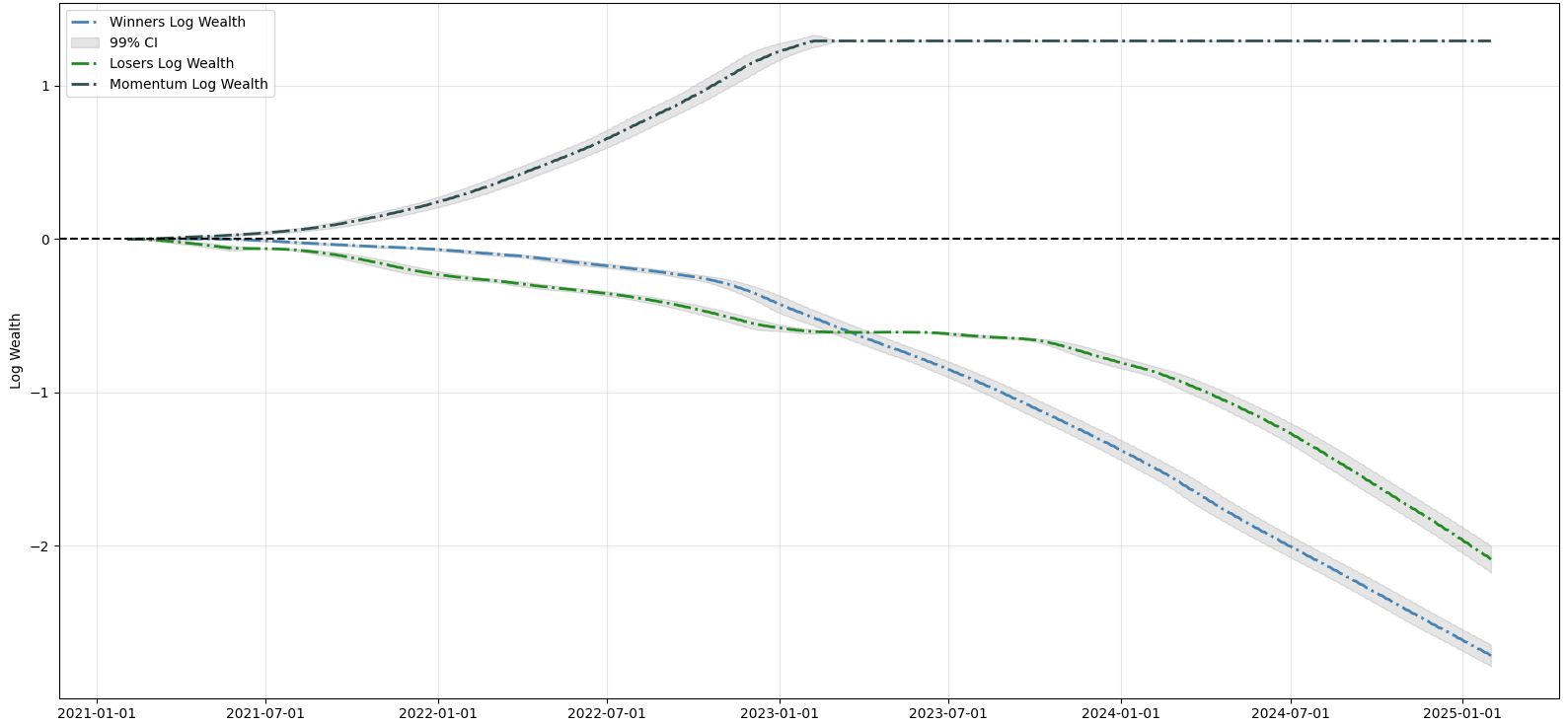}
    \caption{Wealth Accumulation in Portfolios}
    \label{fig:wealth}
\end{figure}

\noindent Figure~\ref{fig:wealth} depicts the log‐wealth evolution under our two‐week formation/two‐week holding strategy.  At every decision date, the CRRA investor computes the optimal weight \(\pi^*(h_t,D_t)\) by balancing the utility gain from higher expected returns against the penalty for risk, where the expected return differential itself jumps whenever \(D_t\), the regime, switches. When the process enters the pro‐ESG state (\(D_t=1\)), the losers portfolio’s risk premium falls by 34.37 basis points and the momentum spread’s premium by 53.54 basis points, signaling that undervalued stocks and the momentum spread are due for a rebound. Since the momentum regime persists with probability \(0.81\), the investor tilts heavily into the momentum spread upon a regime shift and maintains exposure across successive periods, producing a sustained upward slope in the momentum log‐wealth line.

\noindent In contrast, the winners portfolio experiences only a small regime‐driven premium compression (–42 basis points) and, importantly, exits the pro‐ESG state swiftly (\(0.18\)).  Thus, any short‐lived gains in winners are quickly reversed as the model rebalances back to a more conservative stance when the regime reverts. This dynamic is evident in the steep downward path of the winners’ log‐wealth curve following each ESG shock, as the policy rule shifts capital away from overvalued assets once the anticipated regulatory tailwinds abate.

\noindent The losers’ wealth path occupies an intermediate position.  With regime persistence \(p_{11}^{L}=0.72\) and switch‐in probability \(p_{01}^{L}=0.69\), the model repeatedly re‐allocates toward losers in expectation of a sustained pro‐ESG repricing.  The compounded effect of entering the favorable regime, capturing the 34.37 point premium correction, and remaining invested through the regime’s modal duration generates a gradual but persistent rise in the losers’ log‐wealth.  

\noindent Therefore, these regime‐dependent trajectories validate our economic hypothesis: ESG policy shifts create transitory mispricings that a dynamic, state‐aware momentum strategy can systematically exploit. By embedding both the size of the regime effect and the regime‐persistence probabilities into the Bellman recursion, we derive an allocation rule that anticipates repricing events and optimally times exposure to the winners, losers, or momentum spread. The resulting forward‐looking wealth paths, most notably the pronounced outperformance of the momentum curve, thereby reflect both the short‐term volatility of ESG‐driven mispricings and the medium‐term stickiness of policy regimes, offering a compelling demonstration of the value of regime‐switching dynamic programming in ESG‐focused portfolio choice.

\noindent For investors, this implies that a momentum-based approach, by dynamically combining positions in winners and losers, offers a promising strategy to maximize long-term wealth accumulation in ESG portfolios, particularly in an environment characterized by climate-policy shocks. From a policy perspective, our findings underscore the importance of understanding how ESG regulatory shifts dynamically alter risk premiums. This, in turn, can inform the design of legislations and policies aimed at regulating climate-related investments, ensuring that market inefficiencies are addressed in a manner that supports sustainable and efficient capital allocation.

\noindent While our empirical study focuses on U.S. equity and cryptocurrency markets, the modeling framework—combining regime-sensitive momentum construction with tail-aware reward–risk metrics—has potential applicability beyond the U.S. context. Markets with more explicit and enforceable ESG mandates, such as those in the European Union, may offer even stronger test cases due to clearer policy shifts and deeper ESG fund penetration. Although non-U.S. markets are outside the scope of the present analysis, future research could extend this framework to equity universes such as the MSCI Europe ESG Leaders Index or the STOXX Europe 600 ESG. Such an extension would help determine whether the dominance of ESG-loser strategies under pro-ESG regimes is unique to the U.S. setting or indicative of broader cross-market behavioral dynamics associated with ESG sentiment cycles.

\section{Conclusion} \label{conclusion}

\noindent Our study has developed a comprehensive, regime‐aware framework for exploiting momentum in ESG‐related equities. By decomposing the market price of risk into a baseline component and a regime‐dependent adjustment, and by estimating a latent two‐state Markov chain for policy sentiment, we derive a dynamic programming solution that endogenously adjusts portfolio weights in response to both volatility forecasts and expected ESG policy shifts. The first pillar of our identification establishes that a two‐week formation, two‐week holding rebalancing rule maximizes the signal‐to‐noise ratio—across a broad set of reward–risk metrics that satisfy monotonicity, quasi‐concavity, scale invariance, and distributional coherence—thereby yielding the largest expected terminal wealth for a CRRA investor. The second pillar demonstrates that embedding regime‐switching premia and regime‐persistence probabilities in the Bellman recursion produces policy rules that systematically tilt toward undervalued “losers” under pro‐ESG regimes and harvest the ensuing rebound through a long–short momentum spread.  

\noindent \noindent Empirically, we show that the pure losers portfolio outstrips the winners portfolio over long horizons, a counterintuitive outcome driven by the repricing of systematic undervaluation under supportive policy regimes, while the winners–losers momentum strategy delivers the most pronounced wealth gains of all. Across the Russell 3000, Dow Jones 30, and a cryptocurrency universe, the two‐week/two‐week rule consistently outperforms simple buy‐and‐hold benchmarks both in terms of terminal wealth and forecasted risk‐adjusted ratios (Sharpe, CVaR, R–Ratio, STARR). In particular, high‐confidence STARR and R–Ratio measures highlight the momentum spread’s ability to capture extreme upside while capping downside risk, and regime‐aware allocations translate these signals into superior cumulative wealth trajectories.  

\noindent Our contributions are twofold.  Theoretically, we bridge the gap between classical momentum literature and state‐dependent dynamic asset allocation by integrating microfounded reward–risk ratios into a regime‐switching Bellman framework.  Empirically, we deliver a robust, structural methodology for forecasting wealth paths and guiding portfolio rebalancing in the face of ESG policy uncertainty. For practitioners, our results offer precise direction on rebalancing frequency, risk‐management criteria, and regime‐driven tilts. For policymakers, they underscore the powerful influence of climate‐policy regimes on asset valuations and the potential for well‐designed regulatory signals to enhance market efficiency. We invite future research to extend this approach to multi‐state policy chains, alternative utility specifications, and cross‐border ESG regimes, thereby deepening our understanding of trend‐based investing in an era of rapid policy evolution.

\newpage

\appendix

\centering \section*{\textbf{APPENDIX}}

\RaggedRight

\section{Intertemporal Wealth and Portfolio Allocation Rule}

In our model, the asset return is given by

\[
r_t = \lambda(D_t)\,h_t + \sqrt{h_t}\,z_t,\quad z_t\sim N(0,1),
\]

where \(h_t\) is the conditional variance (estimated from an ARFIMA--FIGARCH model) and \(D_t\) is an exogenous ESG policy indicator. In order to capture the evolving effects of ESG regulatory conditions on the risk premium, we model \(D_t\) as a Markov chain with transition probabilities

\[
p_{ij} = \Pr\left(D_{t+1}=j \mid D_t=i\right),
\]

where \(D_t\) takes on a finite set of values (for example, \(D_t=0\) under neutral/anti-ESG conditions and \(D_t=1\) under pro-ESG conditions). We allow the market price of risk to depend on the policy regime via

\[
\lambda(D_t)=\lambda_0+\lambda_1\,D_t.
\]

Thus, when \(D_t=1\) (pro-ESG), the effective market price of risk is \(\lambda_0+\lambda_1\), and when \(D_t=0\) it is \(\lambda_0\).

An investor with CRRA utility 

\[
U(W)=\frac{W^\gamma}{\gamma},\quad \gamma<0,
\]

allocates wealth between a risky asset and a risk-free asset (earning \(r_{rf}\)) according to the wealth dynamics

\[
W_{t+1} = W_t\,\exp\Bigl\{\pi_t\,(r_t-r_{rf})+r_{rf}\Bigr\}.
\]

Owing to the homogeneity of CRRA utility, we conjecture that the value function takes the form

\[
V_t(W,h,D_t)=\frac{W^\gamma}{\gamma}\exp\{J_t(h,D_t)\},
\]

with the terminal condition

\[
J_T(h,D_T)=0 \quad \text{for all } h \text{ and } D_T.
\]

By Bellman’s principle, the investor’s optimization problem is expressed as

\[
V_t(W,h,D_t) = \max_{\pi\in[0,1]}\,E_t\left[V_{t+1}\Bigl(W\,\exp\{ \pi\,(r_t-r_{rf})+r_{rf}\},\,h_{t+1},\,D_{t+1}\Bigr)\right].
\]

Substituting the conjectured form of the value function yields

\[
\frac{W^\gamma}{\gamma}\exp\{J_t(h,D_t)\} = \max_{\pi\in[0,1]}\,E_t\left[\frac{\Bigl(W\,\exp\{ \pi\,(r_t-r_{rf})+r_{rf}\}\Bigr)^\gamma}{\gamma}\exp\{J_{t+1}(h_{t+1},D_{t+1})\right].
\]

Since

\[
\Bigl(W\,\exp\{ \pi\,(r_t-r_{rf})+r_{rf}\}\Bigr)^\gamma = W^\gamma\,\exp\Bigl\{\gamma\Bigl(\pi\,(r_t-r_{rf})+r_{rf}\Bigr)\Bigr\},
\]

dividing both sides by \(W^\gamma/\gamma\) and taking logs gives

\[
J_t(h,D_t) = \max_{\pi\in[0,1]} \ln\,E_t\left[\exp\Bigl\{\gamma\Bigl(\pi\,(r_t-r_{rf})+r_{rf}\Bigr) + J_{t+1}(h_{t+1},D_{t+1})\Bigr\}\right].
\]

With the return process specified as

\[
r_t = (\lambda_0+\lambda_1\,D_t)\,h_t+\sqrt{h_t}\,z_t,
\]

and the state variable \(h\) evolving according to

\[
h_{t+1} = \omega+\beta\,h_t+(\alpha+\ell\,\mathbf{1}\{z_t<0\})h_t\,z_t^2,
\]

the conditional expectation in the Bellman equation must account for the exogenous ESG policy variable \(D_{t+1}\), which is governed by the Markov chain with transition probabilities \(p_{ij}\). Hence, the Bellman equation becomes

\begin{align}
J_t(h,D_t)
&= \gamma\,r_{rf}
+ \max_{\pi\in[0,1]}
  \ln
  \sum_{d'}p_{D_t,d'}\,
  E_z\Bigl[\exp\bigl\{
    \gamma\,\pi\bigl((\lambda_0+\lambda_1\,D_t)h_t+\sqrt{h_t}\,z-r_{rf}\bigr) \nonumber\\
&\quad\quad\quad\quad\;\;
    +\;J_{t+1}\bigl(\omega+\beta\,h_t+(\alpha+\ell\,\mathbf{1}\{z<0\})h_t\,z^2,\;d'\bigr)
  \bigr\}\Bigr]\,.
\end{align}

with \(J_T(h,D_T)=0\) for all \(h\) and \(D_T\).

The optimal portfolio policy is defined as

\[
\pi^*(h,D_t)=\arg\max_{\pi\in[0,1]}\ln\,\sum_{d'} p_{D_t,d'}\,E_z\left[\exp\Bigl\{\gamma\,\pi\Bigl((\lambda_0+\lambda_1\,D_t)h_t+\sqrt{h_t}\,z-r_{rf}\Bigr)+J_{t+1}(h_{t+1},d')\Bigr\}\right],
\]

where 

\[
h_{t+1} = \omega+\beta\,h_t+(\alpha+\ell\,\mathbf{1}\{z_t<0\})h_t\,z_t^2.
\]

Once the function \(J_t(h,D_t)\) is computed by backward induction over a grid of \(h\) values and given the Markov chain evolution of \(D_t\), the investor’s wealth is updated via

\[
W_{t+1} = W_t\,\exp\Bigl\{\pi^*(h_t,D_t)\,(r_t-r_{rf})+r_{rf}\Bigr\},
\]

with

\[
r_t = (\lambda_0+\lambda_1\,D_t)\,h_t+\sqrt{h_t}\,z_t.
\]

\section{ARFIMA (1, $d(m)$,1) - FIGARCH (1, $d(v)$,1)} \label{ARFIMA}

Long-memory effects in the mean, as captured by ARFIMA, indicate that historical returns exert an influence over future values for extended periods. Analogously, when volatility exhibits long memory, as modeled by FIGARCH, episodes of pronounced market fluctuations tend to persist. To incorporate both of these features, we specify the arithmetic returns \(x_{t}\) using the ARFIMA(1, \(d(m)\), 1) process:

\[
\phi(L)\,(1 - L)^{d(m)} x_{t} \;=\; \theta(L)\,\varepsilon_{t},
\]

where \(d(m)\) represents the fractional differencing parameter governing long memory in the mean, and \(\varepsilon_{t}\) is a white noise innovation. For volatility, we employ a FIGARCH(1, \(d(v)\), 1) framework:

\[
\phi(L)\,(1 - L)^{d(v)}\,\varepsilon_{t}^{2} \;=\; \omega \;+\; [1 - \beta(L)]\,\nu_{t},
\]

in which the parameter \(d(v)\) measures the degree of long-memory dependence in volatility and $\nu_{t}$ is the normal innovations term.

\section{Reward-Risk Ratios}

Let \(X\) be a random portfolio return in a given period. We write \(\mathbb{E}[\,\cdot\,]\) for expectation under the physical measure and \(\sigma(\cdot)\) for variance. The four reward-risk ratios, defined by \citet{CheriditoKromer2013}, used to evaluate portfolio performance are as follows:

\begin{enumerate}
    \item \textbf{Sharpe Ratio:} \\ \begin{align*}
        SR(X) = \frac{\mathbb{E}[X]^{+}}{\sigma(X)}
    \end{align*}
    \item \textbf{Stable Tail-Adjusted Return Ratio (STARR):} \\ \begin{align*}
        \text{STARR}_{\gamma}(X) = \frac{\mathbb{E}[X]^{+}}{\text{AVaR}[X]^{+}}
    \end{align*} \\ where $\text{AVaR}_{\gamma}[X] := \gamma^{-1}\int_{0}^{\gamma}\text{VaR}_{u}(X)du$ is the Average-Value-at-Risk at confidence level $\gamma \in (0,1]$.
    \item \textbf{Rachev Ratio (R-Ratio):} \\ 
    \begin{align*}
        RR_{(\beta,\,\gamma)}(X) := \frac{\text{AVaR}_{\beta}[-X]}{\text{AVaR}_{\gamma}[X]} \\ 
    \end{align*} \\ where $\text{AVaR}_{(\beta,\,\gamma)}[X] := \beta^{-1}\int_{0}^{\beta}[max(-F^{-1}_{X}(u),0]^{\gamma}du$.
    \item \textbf{Conditional Value-at-Risk Ratio (CVaR):} \\
    \begin{align*}
        \text{CVaR}_{\gamma}(X) := \frac{\mathbb{E}[X | X<\text{VaR}]^{+}}{\text{VaR}_{\gamma}(X)^{+}}
    \end{align*} \\ where $\text{VaR}_{\gamma}(X) = \text{inf}\{m \in \mathbb{R}:\mathbb{P}[X +m<0] \leq \gamma\}$ is the Value-at-Risk at the confidence level $\gamma \in (0,1]$.
\end{enumerate}

\noindent Denote a generic ratio by \(\alpha(X)\). The four microfounded axioms are:

\begin{enumerate}
\item \emph{Monotonicity:} \quad If \(X\ge Y\) almost surely, then
\[
\alpha(X)\;\ge\;\alpha(Y).
\]
\item \emph{Quasi‐concavity:} \quad For any \(X,Y\) and \(\lambda\in[0,1]\),
\[
\alpha\bigl(\lambda X + (1-\lambda)Y\bigr)
\;\ge\;
\min\{\alpha(X),\,\alpha(Y)\}.
\]
\item \emph{Scale Invariance:} \quad For any scalar \(\gamma>0\),
\[
\alpha(\gamma X)
\;=\;
\alpha(X).
\]
\item \emph{Distribution‐Based:} \quad If \(X\) and \(Y\) have the same distribution under \(\mathbb{P}\), then
\[
\alpha(X)
\;=\;
\alpha(Y).
\]
\end{enumerate}

\section{ Efficient Frontier Analysis}

\begin{figure}[htbp]
    \centering
    \includegraphics[width=0.8\linewidth]{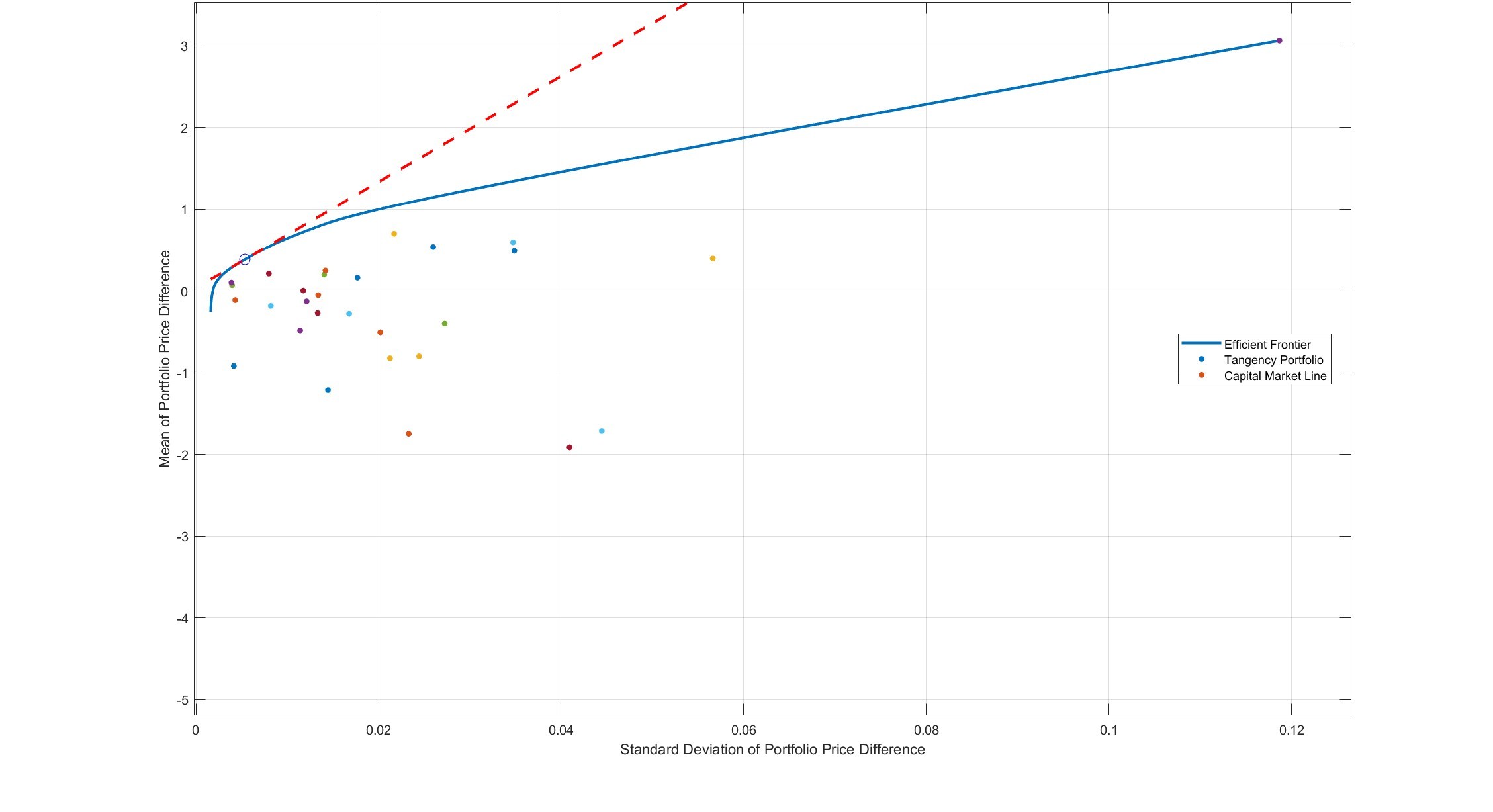}
    \caption{Markowitz based Efficient Frontier}
    \label{fig:mef}
\end{figure}

\begin{figure}[htbp]
    \centering
    \includegraphics[width=1\linewidth]{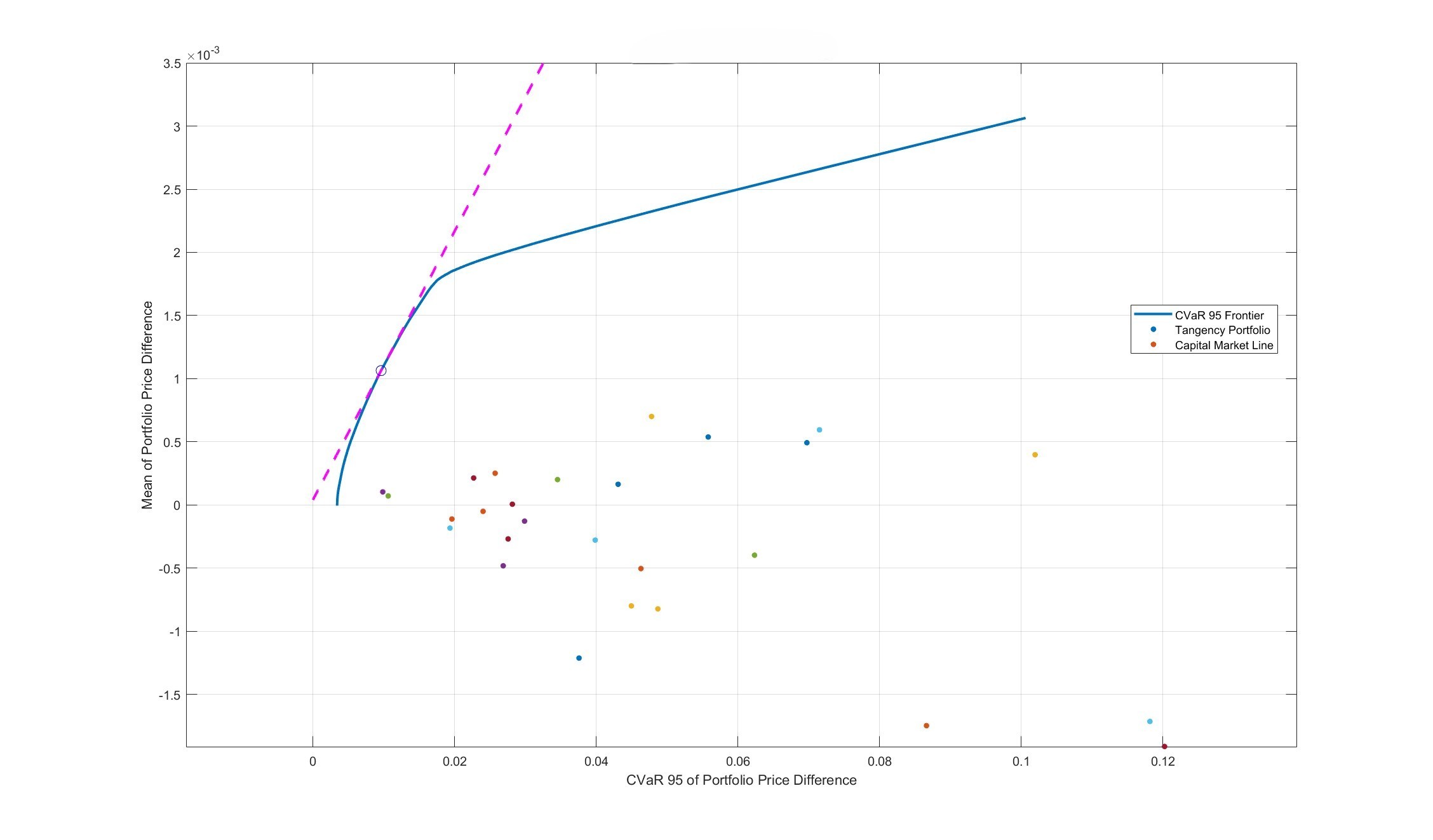}
    \caption{CVaR95 Efficient Frontier}
    \label{fig:cvar95}
\end{figure}

\begin{figure}[htbp]
    \centering
    \includegraphics[width=1\linewidth]{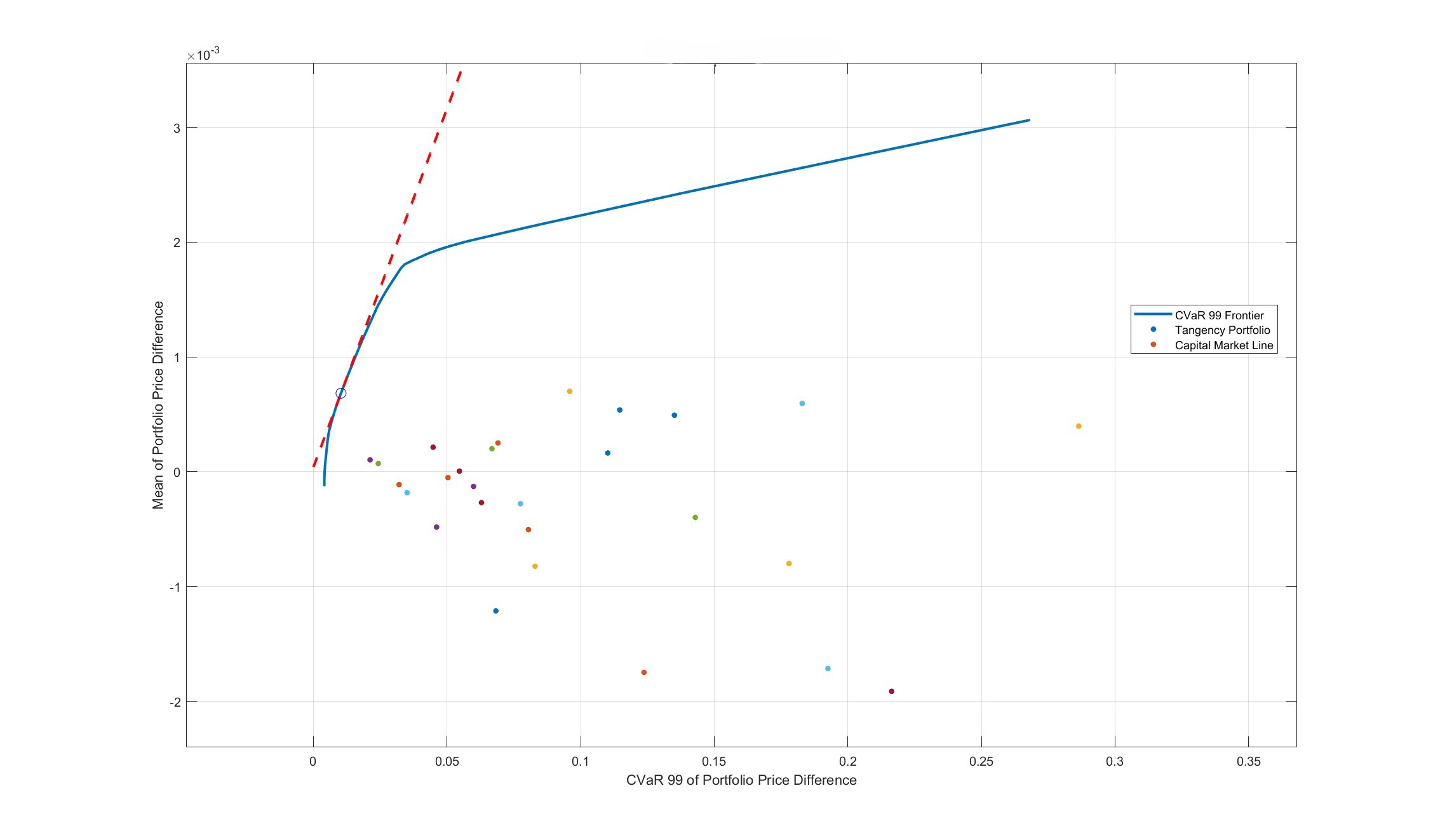}
    \caption{CVaR99 Efficient Frontier}
    \label{fig:cvar99}
\end{figure}

To complement our dynamic, regime‐dependent portfolio results, we also construct three \emph{forward-looking} “efficient frontiers” over the same universe of 30 principal‐component portfolios of the Russell 3000. Each frontier is computed from 10,000 simulated scenarios of forward‐looking returns generated by an ARMA(1,1)–GARCH(1,1) filter with NIG innovations.  In each case we solve

\[
\max_{w\in\mathcal{W}}\;\mathbb{E}\bigl[R(w)\bigr]
\quad\text{s.t.}\quad
\mathcal{R}(w)\;=\;\rho,
\quad\mathbf{1}^\top w = 1,
\]

where \(w\) is the vector of portfolio weights, \(R(w)\) the simulated price‐difference return, and \(\mathcal{R}(\cdot)\) the chosen risk functional.  By varying the risk bound \(\rho\), we trace out an “efficient frontier” of maximum expected return for each risk level.  In the three panels below, we compare:

\begin{enumerate}
  \item \textbf{Mean–Variance (Markowitz) Frontier:}  Here \(\mathcal{R}(w)=\sqrt{\mathrm{Var}[R(w)]}\).  The blue curve in Figure~\ref{fig:mef} plots
  \[
    \max_{w}\;\mathbb{E}[R(w)]
    \quad\text{s.t.}\quad
    \sqrt{\mathrm{Var}[R(w)]} = \sigma,
    \quad\mathbf{1}^\top w = 1.
  \]
  The dashed red line is the Capital Market Line from the risk‐free point, and the gold marker denotes the tangency portfolio, i.e.\ the maximum Sharpe‐ratio allocation.

  \item \textbf{CVaR\,95\% Frontier:}  Here \(\mathcal{R}(w)=\mathrm{CVaR}_{0.95}[R(w)]\), the average loss in the worst 5\% of scenarios.  Figure~\ref{fig:cvar95} shows
  \[
    \max_{w}\;\mathbb{E}[R(w)]
    \quad\text{s.t.}\quad
    \mathrm{CVaR}_{0.95}[R(w)] = \ell,
    \quad\mathbf{1}^\top w = 1.
  \]
  By controlling moderate tail‐risk, this frontier trades off a stricter downside bound against higher mean return at each level of \(\ell\).  

  \item \textbf{CVaR\,99\% Frontier:}  Here \(\mathcal{R}(w)=\mathrm{CVaR}_{0.99}[R(w)]\), the average loss in the worst 1\% of scenarios.  In Figure~\ref{fig:cvar99} we solve
  \[
    \max_{w}\;\mathbb{E}[R(w)]
    \quad\text{s.t.}\quad
    \mathrm{CVaR}_{0.99}[R(w)] = m,
    \quad\mathbf{1}^\top w = 1.
  \]
  This most conservative tail‐risk constraint produces a frontier that is steep at low loss levels but rises sharply once small amounts of extreme‐tail exposure are permitted.
\end{enumerate}

\noindent Across all three risk measures, the tangency portfolios move progressively to the northwest as we tighten tail‐risk (from variance to CVaR 95\% to CVaR 99\%).  In particular, the CVaR\,99\% tangency exhibits the lowest expected shortfall for a given mean return, making it the natural choice for an investor most concerned with extreme climate‐policy shocks.  Together, these frontiers illustrate how a single set of 30 PCA factors can be recombined to optimize for very different definitions of risk, and they underscore the value of our forward‐looking, regime‐aware simulation framework for guiding both mean‐variance and tail‐risk‐sensitive asset allocation.

\appendix

\end{document}